\pdfoutput=1
\documentclass[fleqn]{2022AMSE}
\usepackage{bm}
\usepackage{subfigure}
\usepackage{etoolbox}
\usepackage{cite}
\begin{document}

\ensubject{Fluid Dynamics}

\ArticleType{RESEARCH PAPER}
\Year{2025}

\title{Physics-informed neural networks for hidden boundary detection and flow field reconstruction}

\author[1]{Yongzheng Zhu}{}%
\author[2]{Weizheng Chen}{}
\author[1]{Jian Deng}{}
\author[1]{Xin Bian}{bianx@zju.edu.cn}


\AuthorCitation{Zhu, Chen, Deng, and Bian}

\address[1]{State Key Laboratory of Fluid Power and Mechatronic Systems, Department of Engineering Mechanics, \\Zhejiang University, Hangzhou 310027, P. R. China}
\address[2]{China Ship Scientific Research Center, Wuxi 214082, P. R. China}

\abstract{
Simultaneously detecting hidden solid boundaries and reconstructing flow fields from sparse observations poses a significant inverse challenge in fluid mechanics. 
This study presents a physics-informed neural network~(PINN) framework designed to infer the presence, shape, and motion of static or moving solid boundaries within a flow field. 
By integrating a body fraction parameter into the governing equations, the model enforces no-slip/no-penetration boundary conditions in solid regions while preserving conservation laws of fluid dynamics. 
Using partial flow field data, the method simultaneously reconstructs the unknown flow field and infers the body fraction distribution, thereby revealing solid boundaries. 
The framework is validated across diverse scenarios, including incompressible Navier–Stokes and compressible Euler flows, such as steady flow past a fixed cylinder, an inline oscillating cylinder, and subsonic flow over an airfoil. 
The results demonstrate accurate detection of hidden boundaries, reconstruction of missing flow data, and estimation of trajectories and velocities of a moving body. Further analysis examines the effects of data sparsity, velocity-only measurements, and noise on inference accuracy. The proposed method exhibits robustness and versatility, highlighting its potential for applications when only limited experimental or numerical data are available.
}

\keywords{Hidden moving boundary, Flow reconstruction, Inverse problem, Physics-informed neural networks}

\setlength{\textheight}{23.6cm}
\thispagestyle{empty}

\maketitle
\setlength{\parindent}{1em}

\vspace{-1mm}
\begin{multicols}{2}

\section{Introduction}
\label{sec:Introduction}
\noindent Identifying hidden solid boundaries in flows is a fundamental challenge with broad implications in aerodynamics, biomedical imaging, and marine engineering. 
In aerodynamics,  reconstructing obstacle shapes from flow data enables optimized aerodynamic shape designs and enhanced flow control strategies~\cite{jameson2003aerodynamic, anderson2011ebook}. 
In biomedical applications, such as hemodynamic simulations, boundary reconstruction~(e.g., of aortic valves~\cite{aguayo2021distributed}) is critical for diagnosing vascular diseases and analyzing blood flow dynamics~\cite{nolte2022inverse, partl2023reconstruction}. 
Similarly, in marine engineering, inferring submerged structures or underwater obstacles from flow measurement is essential for ship hydrodynamics and offshore infrastructure monitoring~\cite{zhang2018reconstruction, rabago2024detecting}.
However,  practical scenarios often lack direct boundary observations due to occlusions, sparse sensor coverage, or unobserved object motion.
Consequently, inferring solid boundaries-including their presence, shape, and motion--while simultaneously reconstructing the surrounding flow field from sparse measurement poses a highly challenging inverse problem.
This task is further complicated by the underlying physics, governed by the Navier-Stokes~(NS) equations for incompressible flows or the Euler equations for compressible flows, which couple boundary conditions with fluid dynamics.
Addressing this problem is vital for applications ranging from flow control and object tracking to inverse fluid dynamics analysis.

Traditional methods for inferring hidden boundaries include data assimilation techniques~\cite{tiago2014patient}, shape optimization approaches~\cite{agoshkov2006mathematical, manzoni2018adjoint}, and immersed resistance methods~\cite{aguayo2021distributed, partl2023reconstruction}. 
While effective in certain cases, these approaches face significant limitations. 
Data assimilation integrates observational data with numerical models to enhance predictions but typically relies on high-quality, full-field velocity measurements for reliable reconstructions~\cite{tiago2014patient}. In practice, sparse or noisy data can severely degrade its accuracy.
Shape optimization (e.g. adjoint-based methods~\cite{agoshkov2006mathematical, manzoni2018adjoint}) iteratively adjusts boundary geometry to minimize discrepancies between simulations and target solutions. However, these techniques are computationally expensive and highly sensitive to measurement uncertainties.
Immersed resistance methods, such as those that reconstruct aortic valve shapes from synthetic magnetic resonance imaging data~\cite{aguayo2021distributed}, represent obstacles via resistance functions in the steady NS equations. While useful, their applicability is often restricted to specific flow regimes or boundary types.

Recent advances in deep learning have revolutionized fluid mechanics,
with techniques such as data-driven modeling, physics-informed constraints, variational formulations, operator learning, and discrete loss optimization enabling accurate approximations of complex flow behaviors~\cite{Brunton2016, sirignano2018dgm, E2018, raissi2019physics, li2021fourier, lu2021learning, karnakov2024solving}.
Among these, physics-informed neural networks~(PINNs)~\cite{raissi2019physics, karniadakis2021physics} have emerged as a particularly powerful framework by embedding governing physical laws directly into neural networks. This integration allows PINNs to provide generalizable and data-efficient solutions for both forward and inverse problems involving partial differential equations~(PDEs)~\cite{cai2021review}.
PINNs are especially promising for inferring hidden boundaries, as they naturally incorporate physical constraints and can effectively leverage sparse flow measurements. Their versatility has led to widespread adoption across fluid mechanics, including applications in compressible flows, heat transfer, multiphase flows, moving boundaries, and more~\cite{raissi2019deep, mao2020physics, cai2021physics, jin2021nsfnets, qiu2022physics, zhu2024physics, zhu2024unstructured}.
A key strength of PINNs lies in their flexibility for inverse problem, such as estimating unknown NS parameters~\cite{raissi2019physics}, reconstructing flow fields from sparse data~\cite{raissi2020hidden}, and achieving super-resolution flow predictions~\cite{fathi2020super, gao2021super}.
Although pioneering studies have successfully reconstructed velocity and pressure fields using PINNs~\cite{raissi2019deep, jin2021nsfnets, zhu2024physics,  cai2021flow}, the simultaneous inference of hidden or moving solid boundaries-critical for defining flow domains-remains underexplored. 
Building on the penalization method proposed in Ref.~\cite{karnakov2024solving}, which incorporates obstacles into governing equations via optimization, we extend this approach within the PINNs framework to enable robust boundary shape inference, particularly for dynamic boundaries. This advancement addresses a key gap in current research.

In this work, we present a novel PINNs-based framework capable of simultaneously inferring hidden stationary/moving boundaries and reconstructing surrounding flow fields from sparse observations.
A key innovation of our approach lies in the introduction of a body fraction parameter into the governing equations, which explicitly distinguishes between fluid and solid regions. This formulation enables the model to automatically enforce domain-appropriate physics: conservation laws in fluid regions and no-slip/no-penetration boundary conditions in solid regions. 
By training the network with limited flow measurements, our method infers the spatial distribution of the body fraction parameter, thereby reconstructing both stationary and moving boundaries.
Notably, for moving boundaries, the framework estimates not only their instantaneous shapes but also their motion characteristics (trajectories and velocities). 
Additionally, the model recovers physically consistent flow fields in unobserved regions, effectively solving the inverse problem while respecting underlying fluid dynamics.
We demonstrate the versatility of our approach across both incompressible NS and compressible Euler flows through three benchmark problems:
Steady flow around a fixed cylinder; transient flow induced by an oscillating cylinder;
subsonic flow over an airfoil. 
In all cases,  the framework successfully reconstructs (i) hidden boundaries (stationary and moving), (ii) complete velocity/pressure fields, and (iii) for moving objects, their kinematics. Comprehensive numerical experiments validate the robustness of our method under spare/noisy measurements, highlighting its potential for real-world applications when boundary data is inaccessible.

The remainder of this paper is structured as follows.
Section~\ref{sec:Overview of PINNs} provides a brief introduction to PINNs.
Section~\ref{sec:PINNs for inferring hidden boundaries} details our proposed methodology, including the integration of the body fraction parameter into the governing equations and the development of specialized PINN architectures for stationary and moving boundary inference.
In Section~\ref{sec:Numerical results}, we demonstrate our framework's effectiveness through three representative test cases, showcasing its ability to reconstruct both hidden boundaries and complete flow fields from limited measurements.
Finally, section~\ref{sec:Conclusions} summarizes our key findings and discusses potential applications.



\section{Physics-informed neural networks}
\label{sec:Overview of PINNs}
We briefly review physics-informed neural networks~(PINNs) for solving partial differential equations~(PDEs)~\cite{raissi2019physics}.
In general, we consider the parametric PDEs for the solution $u({\bm x})$ with ${\bm x}=\left(x_1, x_2, \ldots, x_d\right)$ defined on a domain $\Omega \subset \Bbb R^d$, in a form expressed as follows:
\begin{equation}
f\left( {{\bm x};\frac{{\partial u}}{{\partial {x_1}}}, \ldots ,\frac{{\partial u}}{{\partial {x_d}}};\frac{{{\partial ^2}u}}{{\partial {x_1}\partial {x_1}}}, \ldots ,\frac{{{\partial ^2}u}}{{\partial {x_1}\partial {x_d}}}; \ldots ; {\lambda}} \right) = 0,
\label{eq:pdes}
\end{equation}
with appropriate boundary and initial conditions:
\begin{eqnarray}
&& {\cal B}(u,{\bm x}) = 0,\quad {\bm x} \in \partial \Omega,
\\
&& {\cal I}(u,{\bm x}) = 0,\quad {\bm x} \in {{\Gamma}_i}.
\end{eqnarray}
Here $d$ is the dimension of the problem and $\lambda=\left(\lambda_1, \lambda_2, \ldots\right)$ are the parameters for combining each component.
The boundary operator ${\cal B}(\cdot)$ defines the boundary conditions on $\partial \Omega$, which may be Dirichlet, Neumann, Robin, or periodic. 
Similarly, the initial operator ${\cal I}(\cdot)$ specifies the initial conditions, with ${\Gamma}_i$ representing the spatial domain of $\bm{x}$ at the initial snapshot.

In the framework of PINNs, a deep neural network is constructed to serve as a highly nonlinear function $\hat{u}(\bm{x};\theta)$ to approximate the unknown solution $u$ of PDEs, where $\theta$ represents all
trainable parameters of the network~(e.g., weights and biases).
This neural network consists of an input layer, multiple hidden layers, and an output layer, where each layer contains neurons with associated weights, biases, and a nonlinear activation function $\sigma(\cdot)$.
Given input $\bm{x}$, the implicit variable at the $l$-th hidden layer, ${\bm y}^{(l)}$, a deep neural network with $L$ layers can be formulated as:
\begin{equation}
\left\{
\begin{aligned}
{\bm y}^{(l)}& = {\bm x}, \quad l=0, \\
{\bm y}^{(l)}& = \sigma\left(\bm{W}^{(l)} {\bm y}^{(l-1)}+\bm{b}^{(l)}\right), \quad 1 \leq l \leq L-1, \\
{\bm y}^{(l)}& = \bm{W}^{(l)} {\bm y}^{(l-1)}+\bm{b}^{(l)}, \quad l=L.
\end{aligned}
\right.
\end{equation}
Here $\bm{W}^{(l)}$ and $\bm{b}^{(l)}$ represent the weight matrix and bias vector, respectively, which are trainable parameters at the $l$-th layer of the neural network.

To ensure that the neural network $\hat{u}(\bm{x};\theta)$ adheres to the physical constraints imposed by the PDEs and boundary/initial conditions, we enforce these conditions at a set of scattered points, which are either randomly distributed or grid-aligned within the domain. 
The PDE residuals associated with Eq.~(\ref{eq:pdes}) can be defined as $\mathcal{R}(\hat{u}, \bm{x})$.
A PINNs model is then trained by minimizing a composite loss function, which integrates the residuals from the PDEs, boundary conditions, initial conditions, and available labeled data:
\begin{equation}
\begin{aligned}
{\cal L}({\bm{\theta}};{\cal T}) = {{\cal L}_f}\left({\cal T}_f \right) + {{\cal L}_{bc}}\left({\cal T}_{bc} \right) + {{\cal L}_{ic}}\left( {\cal T}_{ic} \right) + {{\cal L}_{d}}\left({\cal T}_{d} \right),
\label{eq:totalloss}
\end{aligned}
\end{equation}
where
\begin{eqnarray}
&&\mathcal{L}_f\left(\bm{\theta} ; \mathcal{T}_f\right)=\frac{1}{\left|\mathcal{T}_f\right|} \sum_{\bm{x} \in \mathcal{T}_f}|\mathcal{R}(\hat{u}, \bm{x})|^2, 
\\
&&\mathcal{L}_{bc}\left(\bm{\theta} ; \mathcal{T}_{bc}\right)=\frac{1}{\left|\mathcal{T}_{bc}\right|} \sum_{\bm{x} \in \mathcal{T}_{bc}}|\mathcal{B}(\hat{u}, \bm{x})|^2,
\\
&&\mathcal{L}_{ic}\left(\bm{\theta} ; \mathcal{T}_{ic}\right)=\frac{1}{\left|\mathcal{T}_{ic}\right|} \sum_{\bm{x} \in \mathcal{T}_{ic}}|\mathcal{I}(\hat{u}, \bm{x})|^2,
\\
&&\mathcal{L}_{d}\left(\bm{\theta}; \mathcal{T}_{d}\right)=\frac{1}{\left|\mathcal{T}_{d}\right|} \sum_{\bm{x} \in \mathcal{T}_{d}}|\hat{u}(\bm{x})-u(\bm{x})|^2.
\end{eqnarray}
Here, ${\cal T}_f$, ${\cal T}_{bc}$, ${\cal T}_{ic}$ and ${\cal T}_{d}$ denote the respective sets of training points located within the domain or on the boundary.
Since the output of the neural network $\hat u$ serves as the approximate solution, each term in the PDEs, such as the partial derivative $\partial \hat{u}/\partial x_1$ or the second-order partial derivative $\partial^2 \hat{u}/\partial x_1^2$, can be efficiently computed by differentiating $\hat u$ with respect to $\bm{x}$ one or more times using automatic differentiation~(AD)~\cite{rall1981automatic, griewank2008evaluating}.
The training stage involves optimizing $\bm{\theta}$ by minimizing the loss function $\mathcal{L} (\boldsymbol{\theta} ; \mathcal{T})$ in Eq.~(\ref{eq:totalloss}). 
Given that $\mathcal{L} (\boldsymbol{\theta} ; \mathcal{T})$ is highly nonconvex and nonlinear for $\bm{\theta}$, gradient-based optimizers such as Adam~\cite{kingma2014adam} and L-BFGS~\cite{liu1989limited}, are usually combined for an effective minimization.

One of the key strengths of PINNs lies in the flexibility of their composite loss function, which enables seamless integration of partial labeled data with physical laws. This adaptability makes PINNs particularly powerful for solving inverse problems, where limited observations can be effectively combined with governing equations to infer unknown parameters or hidden dynamics.

\section{PINNs for inferring hidden boundaries}
\label{sec:PINNs for inferring hidden boundaries}
We shall further introduce the concept of embedding a body fraction parameter into the governing equations to enforce Dirichlet boundary penalty terms, enabling differentiation of solid regions from fluid regions.  
We then present two architectures of neural network designed for inferring both stationary and moving boundaries while integrating PDE constraints with available data. 
Finally, we detail the loss functions and key technical aspects of the proposed framework.
\subsection{Embedding a body fraction parameter in governing equations}
\subsubsection{Navier-Stokes equation}
The Navier-Stokes~(NS) equations governing an viscous incompressible flow in two-dimensional Cartesian coordinates are expressed in the following vector form:
\begin{eqnarray}
\label{eq:NS equation}
\begin{aligned}
    \nabla  \cdot {\bm u}& = 0,
    \\
    \frac{{\partial {\bm u}}}{{\partial t}} + ({\bm u} \cdot \nabla ){\bm u}& =  - \nabla p + \frac{1}{{Re}}{\nabla ^2}{\bm u},
\end{aligned}
\end{eqnarray}
where $\bm {x}=(x, y)\in \Omega \subset \mathbb{R}^2$, the velocity vector \({\bm u} = (u, v)\) consists of the velocity components \(u\) and \(v\) in the \(x\)- and \(y\)-directions, respectively, and \(p\) is the pressure. \(Re\) represents the Reynolds number defined as \(Re = UL/\nu\) where \(U\) is the characteristic velocity, \(L\) is the characteristic length, and \(\nu\) is the kinematic viscosity. \(\nabla = \left( \partial / \partial x, \partial / \partial y \right)\) is the gradient operator and \(\nabla^2 = \partial^2 / \partial x^2 + \partial^2 / \partial y^2\) is the Laplace operator. 
A typical forward problem is solved by incorporating one or more boundary conditions for velocity and/or pressure as follows:
\begin{eqnarray}
\label{eq:Dirichlet_velocity_bcs}
\begin{aligned}
{\bm u} = {{\bm u}_\Gamma }({\bm x}),\quad {\bm x} \in {\Gamma _D},
\\
p = {p_\Gamma }({\bm x}),\quad {\bm x} \in {\Gamma _D},
\\
\frac{{\partial {\bm u}}}{{\partial {\bm n}}} = 0,\quad {\bm x} \in {\Gamma _N},
\\
\frac{{\partial p}}{{\partial {\bm n}}} = 0,\quad {\bm x} \in {\Gamma _N},
\end{aligned}
\end{eqnarray}
where ${\Gamma _D}$ and ${\Gamma _N}$ indicate the Dirichlet and Neumann boundaries, respectively; ${\bm u}_\Gamma$ and $p_\Gamma$ are the corresponding boundary constraints. ${\bm n}$ is the normal vector at the corresponding boundary.

However, in inverse problems, boundary conditions such as those in Eq.~(\ref{eq:Dirichlet_velocity_bcs}) are usually infeasible. 
Instead, only a limited number of uniformly or randomly distributed measurement points are provided. Moreover, the available data at these points are often incomplete, containing either velocity or pressure values but not both. 
Reconstructing the surrounding flow field and even inferring hidden boundaries within the flow using sparse and limited data still presents a significant challenge.  
Assuming the presence of a stationary or translating object in the flow field, without considering its rotation, our objective is to infer its shape and motion using only limited flow field data.
Inspired by the penalization method adopted in the governing equations~\cite{angot1999penalization, karnakov2024solving}, we propose a generalized NS equations that incorporates Dirichlet boundary penalty terms, enabling the inference of both stationary and moving boundaries, expressed as:
\begin{eqnarray}
\label{eq:penalized NS equation}
\begin{aligned}
    \nabla  \cdot {\bm u} = 0,
    \\
    \left( {1 - \phi } \right)\left( \frac{\partial u}{\partial t} + ({\bm u} \cdot \nabla) u + \frac{\partial p}{\partial x} - \frac{1}{Re} \nabla^2 u \right)
    \\
    + \alpha  \cdot \phi  \cdot \left( {u - U} \right) = 0,
    \\
    \left( {1 - \phi } \right)\left( \frac{\partial v}{\partial t} + ({\bm u} \cdot \nabla) v + \frac{\partial p}{\partial y} - \frac{1}{Re} \nabla^2 v \right)
    \\
    + \alpha  \cdot \phi  \cdot \left( {v - V} \right) = 0,
\end{aligned}
\end{eqnarray}
where \( \phi \) represents the body fraction parameter, and \( \alpha \) is the penalization parameter. The parameter \( \phi \) distinguishes between fluid (\(\phi = 0\)) and solid (\(\phi = 1\)) regions, allowing the model to enforce appropriate physical constraints, that is, applying standard conservation laws in the fluid domain and Dirichlet velocity boundary conditions in the solid region.  
Subsequently, the shape of the solid body can be constructed by the contour of \( \phi = 1 \). 
The velocity components of a moving solid body denoted by \( U \) and \( V \)
can also be inferred. 

If we know that the solid body is stationary in advance~(but we still do not know
its position and shape),
Eq.~(\ref{eq:penalized NS equation}) can be simplified as:
\begin{eqnarray}
\label{eq:steady penalized NS}
\begin{aligned}
    \nabla  \cdot {\bm u} = 0,
    \\
    (1-\phi)\cdot(({\bm u} \cdot \nabla ){\bm u}  + \nabla p - \frac{1}{{Re}}{\nabla ^2}{\bm u}) + \alpha\cdot\phi\cdot{\bm u} = 0.
\end{aligned}
\end{eqnarray}
Therefore, the architecture of the neural network can also be simplified
as described in Sec.~\ref{subsec_pinns_archi}.

If the body moves over time, \( U(t) \) and \( V(t) \) become unknown time-dependent functions.
Therefore, the flow field on the surface of the body should satisfy the corresponding velocity boundary conditions at each time instance. Based on the Dirichlet boundary condition in Eq.~(\ref{eq:Dirichlet_velocity_bcs}), we have \( u = U(t) \) or/and \( v = V(t) \).
For an inverse problem involving flows with moving boundaries, the unknown boundary information includes not only the shape of the hidden body but also its velocity. After the body's velocity and initial position are inferred, its motion trajectory can be obtained through simple numerical integration.

\subsubsection{Euler equation}
The Euler equations governing an inviscid compressible flow are expressed in the following conservative form:
\begin{equation}
\label{eq:Euler}
\begin{split}
    \frac{\partial \bm{U}}{\partial t}+{\nabla} \cdot {\bm{f(U)}}=0,\quad \bm{x}\in \Omega \subset \mathbb{R}^2,
\end{split}
\end{equation}
where $\bm{U}$ is the vector of the conservative quantities with density, $\bm{f}$ denotes the solution flux matrix. In two-dimensional Cartesian coordinates, they are defined as:
\begin{eqnarray}
    \bm{U}=\left[\begin{array}{c}
    \rho \\
    \rho u \\
    \rho v \\
    \rho E
    \end{array}\right],\quad 
    \bm{f}=(G_1, G_2),
\end{eqnarray}    
\text {with} 
\begin{equation}
    G_1(\bm{U})=\left[\begin{array}{c}
    \rho u \\
    \rho u^2 + p \\
    \rho u v \\
    u(\rho E +p)
    \end{array}\right],
    G_2(\bm{U})=\left[\begin{array}{c}
    \rho v \\
    \rho u v \\
    \rho v^2 + p\\
    v(\rho E +p)
    \end{array}\right] \text {, }
\end{equation}
where \( (u, v) \) represent the velocity components, \( p \) denotes the pressure, \( \rho \) is the density, and \( E \) corresponds to the total specific energy. To fully define the Euler equations, an equation of state is necessary to establish the relationship between pressure \( p \) and energy \( E \). Specifically, we adopt the equation of state for an ideal gas, given by:
\begin{equation}
p=(\gamma-1)\left(\rho E-\frac{1}{2} \rho\|\boldsymbol{u}\|^2\right).
\end{equation}
Here, \( \gamma = 1.4 \) represents the specific heat ratio of air, and the velocity vector is given by \( \bm{u} = (u, v) \). The Mach number, defined as \( M = \|\bm{u}\| / a \), represents the ratio of the flow velocity to the speed of sound, where the sound speed is expressed as \( a = \sqrt{\gamma p / \rho} \).

In this study, we focus on the steady-state solution, which is particularly relevant in aerodynamic applications.  
To nondimensionalize the flow properties, we use the speed of sound $a_0$ and density $\rho_0$ at the stagnation point, leading to the following dimensionless variables:  
\begin{equation}
    u^{\prime}=\frac{u}{a_0}, v^{\prime}=\frac{v}{a_0}, \rho^{\prime}=\frac{\rho}{\rho_0}, p^{\prime}=\frac{p}{\rho_0 \cdot a_0^2}.
\end{equation}
In this dimensionless formulation, the steady Euler equations retain the same form as the dimensional ones of Eq. (\ref{eq:Euler}), with the time derivative omitted and prime symbols (\(^\prime\)) removed for clarity.   
Note that we adopt this dimensionless form of the Euler equations for both training and prediction in PINNs.

For inviscid compressible flows, slip velocity boundary conditions are typically applied at solid walls. Unlike no-slip boundaries, where a fixed velocity can be directly imposed, the velocity on a slip boundary is entirely determined by the nearby flow, making it seemingly impractical to enforce a spatially independent velocity constraint. However, since slip walls still satisfy the no-penetration condition, we argue that for a stationary boundary, a fixed velocity constraint can still be applied to ensure that no flow exists inside the solid region.  
Here, we enforce zero velocity to impose the no-penetration condition at the wall.
Similar to the NS equations, we introduce a velocity penalty term into the momentum equations of the Euler equations for the first time. 
For steady flows, the penalized Euler equations can be expressed as:
\begin{equation}
\label{eq:penalized Euler}
\begin{aligned}
    \frac{\partial (\rho u)}{\partial x}+\frac{\partial (\rho v)}{\partial y}=0,
    \\
    (1-\phi)\cdot (\frac{\partial (\rho u^2 +p)}{\partial x}+\frac{\partial (\rho u v)}{\partial y}) + \alpha\cdot\phi\cdot u=0,
    \\
    (1-\phi)\cdot (\frac{\partial (\rho u v)}{\partial x}+\frac{\partial (\rho v^2+p)}{\partial y}) + \alpha\cdot\phi\cdot v=0,
    \\
    \frac{\partial (u(\rho E +p))}{\partial x}+\frac{\partial (v(\rho E +p))}{\partial y}=0,
\end{aligned}
\end{equation}
where \( \phi \) represents the body fraction parameter, and \( \alpha \) is the penalization parameter. 
Also, the shape of the body can be determined by \( \phi \), which is assigned \( \phi = 1 \) within the body region and \( \phi = 0 \) in the surrounding fluid. 

\subsection{PINNs for stationary and moving boundaries}
\label{subsec_pinns_archi}
\begin{figure*}[htb!]
    \centering
    \includegraphics[width=1.0\linewidth]{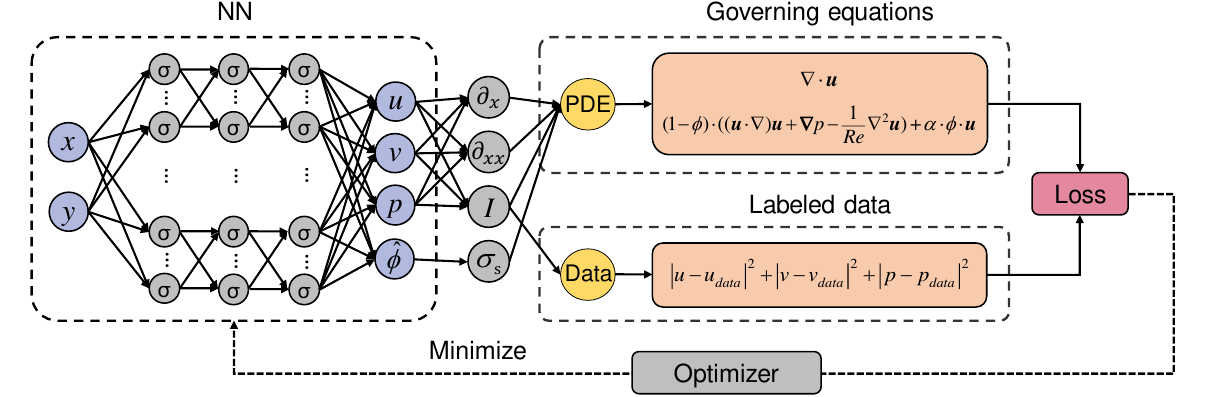}
    \caption{Schematic of PINNs for inferring the shape of a hidden {\it stationary} boundary. 
    For the NS equations, the network models the flow field by predicting velocity $(u, v)$, pressure $p$, and body fraction $\phi$, taking spatial coordinates $(x, y)$ as inputs. 
    For the Euler equations, the only difference is that the network also outputs density $\rho$. 
    The loss functions comprise two components: a physical equation part and a known labeled data part.}
    \label{fig:NN_for_static_bc}
\end{figure*}
\begin{figure*}[htb!]
    \centering
    \includegraphics[width=1.0\linewidth]{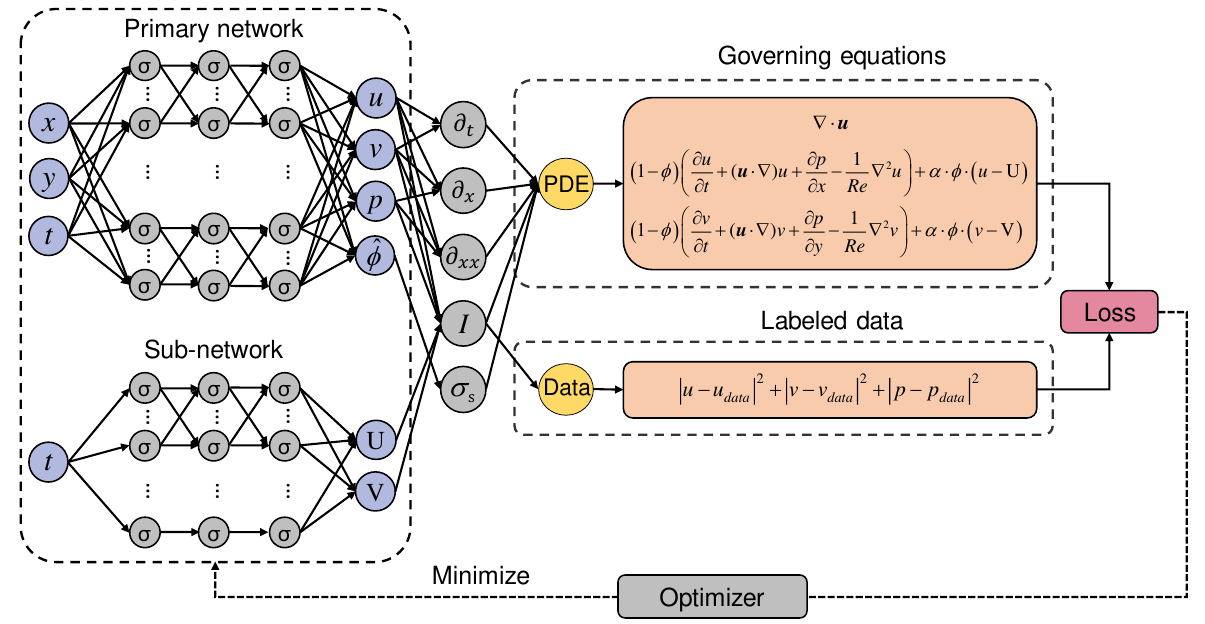}
    \caption{Schematic of PINNs for inferring the shape, velocity, and trajectory of a hidden {\it moving} boundary. 
    The network architecture consists of two components: a primary network and a sub-network. 
    The primary network models the flow field by predicting velocity $(u, v)$, pressure $p$, and body fraction $\phi$, taking spatial coordinates $(x, y)$ and time $t$ as inputs. 
    The sub-network captures the velocity of the moving boundary, using time $t$ as input and outputting the boundary velocity components $U$ and $V$. Both networks optimized simultaneously. 
    The loss functions comprise two components: a physical equation part and a known labeled data part.}
    \label{fig:NNs_for_moving_bc}
\end{figure*}

Here, we will present the technical details of two PINN architectures for solving inverse problems involving both stationary and moving boundaries, along with a detailed illustration of each component of the composite loss function.
For the inverse problems under consideration, the only available information consists of the governing physical equations of the flow and a limited set of labeled data points,
that is, without boundary or initial conditions.
Based on the principles of PINNs outlined in in Sec.~\ref{sec:Overview of PINNs}, the loss functions comprise two components: a physical equation part and a known labeled data part, formulated as follows:
\begin{equation}
\begin{aligned}
{\cal L}({\bf{\theta }}) = {w_f}{{\cal L}_f}({\bf{\theta }}) + {w_d}{{\cal L}_d}({\bf{\theta }}).
\label{eq:totalloss2}
\end{aligned}
\end{equation}
where ${\cal L}_f$ and ${\cal L}_d$ denote the PDEs loss and the labeled data loss, respectively. $\theta$ represents all trainable parameters~(e.g., weights and biases) of the neural network. $w_f$ and $w_d$ are the weights of each loss term. 
Each component of the loss function is also illustrated in Fig.~\ref{fig:NN_for_static_bc} and Fig.~\ref{fig:NNs_for_moving_bc} for a clear overview.

Fig.~\ref{fig:NN_for_static_bc} illustrates a schematic of PINNs for inferring the shape of a hidden {\it stationary} boundary. 
For the NS equations, the neural network takes spatial coordinates $(x, y)$ as inputs and outputs $(u, v, p, \hat{\phi})$ to predict velocity, pressure, and body fraction.
In the case of inviscid compressible flow governed by the Euler equations, the network additionally outputs density $\rho$, making the output $(u, v, p, \rho, \hat{\phi})$.
To ensure that $\hat{\phi}$ accurately represents the body fraction, it is transformed using a sigmoid-like activation function:  
\begin{equation}
\begin{aligned}
\phi = \frac{1}{1 + \exp(-10 \cdot \hat{\phi})}
\label{eq:transform_phi}
\end{aligned}
\end{equation}
This mapping forces \( \phi \) to take values strictly within [0,1], creating a sharp distinction between 0 and 1 due to the narrow transition zone. 
As a result, \( \phi \) can serve as a reliable indicator of the body fraction and effectively distinguish between solid and fluid regions.
The processed \( \phi \) is then embedded into the governing equations to construct the loss function.
Taking the penalized Euler equations in Eq.~(\ref{eq:penalized Euler}) as an example, the four components $f_i$ correspond to the mass, penalized momentum, and energy conservation equations. 
The definitions of all loss terms are formulated as mean squared errors (MSE), and specifically, they are given as  
\begin{equation}  
\begin{aligned}  
{{\cal L}_f} = \frac{1}{{{N_f}}}\sum\limits_{i = 1}^4\sum\limits_{n = 1}^{N_{f}}{\left| f_i(\bm x^n) \right|^2},  
\end{aligned}  
\end{equation}  
\begin{equation}
\begin{aligned}
{{\cal L}_{d}} = \frac{1}{{N_{d}}}\sum\limits_{n = 1}^{N_{d}} ( {\left| {u({\bm x}_u^n) - u_{d}^n} \right|}^2 
+ {\left| {v({\bm x}_v^n) - v_{d}^n} \right|}^2\\
+ {\left| {p({\bm x}_p^n) - p_{d}^n} \right|}^2
+ {\left| {\rho({\bm x}_\rho^n) - \rho_{d}^n} \right|}^2),
\end{aligned}
\end{equation}
where $f_i(\bm{x}^n)$ represents the residual of the $i$-th equation at training point $\bm{x}^n$, $N_f$ is the total number of training points used for enforcing the physics-informed constraints, and $N_{d}$ denotes the number of labeled data points.
The derivatives of $u$, $v$, $p$ and $\rho$ with respect to the inputs are calculated using the automatic differentiation. 

Fig.~\ref{fig:NNs_for_moving_bc} illustrates a schematic of PINNs for inferring the shape, velocity, and trajectory of a hidden {\it moving} boundary. The network architecture consists of two components: a primary network and a sub-network. 
The primary network models the flow field by predicting velocity $(u, v)$, pressure $p$, and body fraction $\hat{\phi}$, taking spatial coordinates $(x, y)$ and time $t$ as inputs. 
Similarly, $\hat{\phi}$ is transformed using the mapping in Eq.~(\ref{eq:transform_phi}) to obtain \(\phi\), which is then embedded into the equations to construct the loss function.
The sub-network captures the velocity of the moving boundary, using time $t$ as input and outputting the boundary velocity components $U$ and $V$. 
Both networks are optimized simultaneously. 
The rationale for employing two separate networks stems from the fact that the velocity of the moving boundary depends solely on temporal parameter $t$. To circumvent potential correlations with spatial coordinates and avoid introducing unforeseen influences, we deliberately exclude the velocity of the solid body from the primary network's output. Instead, we implement an independent network dedicated to approximating this specific physical quantity.
Since the sub-network is not intended to approximate highly complex functions, its architecture can be designed with fewer layers compared to the primary network, thereby maintaining computational efficiency.
The loss functions comprise two components: a physical equation part and a known labeled data part.
Taking the penalized NS equations in Eq.~(\ref{eq:penalized NS equation}) as a demonstration, the three components $f_i$ correspond to the continuity and penalized momentum equations. 
The definitions of all loss components are formulated as follows:  
\begin{equation}  
\begin{aligned}  
{{\cal L}_f} = \frac{1}{{{N_f}}}\sum\limits_{i = 1}^3\sum\limits_{n = 1}^{N_{f}}{\left| f_i(\bm x^n, t^n) \right|^2},  
\end{aligned}  
\end{equation}  

\begin{equation}  
\begin{aligned}  
{{\cal L}_{d}} = \frac{1}{{N_{d}}}\sum\limits_{n = 1}^{N_{d}} \left( {\left| {u({\bm x}_u^n, t_u^n) - u_{d}^n} \right|}^2  
+ {\left| {v({\bm x}_v^n, t_v^n) - v_{d}^n} \right|}^2 \right),  
\end{aligned}  
\end{equation}  
where $N_f$ and $N_{d}$ represent the number of training points used for the physics-informed loss and the known data loss, respectively.
In this work, the penalization parameter \(\alpha\) in the governing equations is set to 1 by default, unless stated otherwise.

The architecture of the neural network, including the number of hidden layers and neurons per layer, is typically chosen to suit the specific requirements of the problem. After conducting sensitivity tests through trial and error, we set the network to consist of 8 hidden layers, each containing 40 neurons.  
For the sub-network in Fig.~\ref{fig:NNs_for_moving_bc}, we employ 2 hidden layers and each layer contains 20 neurons. 
For the activation function $\sigma(\cdot)$, we employ the smooth and differentiable hyperbolic tangent function \textit{tanh}. 
The network parameters, including weights and biases denoted as $\theta$, are initialized using the Glorot scheme~\cite{glorot2010understanding} to ensure stable training.  
To optimize the total loss function, we use a combination of the Adam optimizer~\cite{kingma2014adam}, which adapts the learning rate dynamically, and the L-BFGS optimizer~\cite{liu1989limited}, a Quasi-Newton method that enhances convergence. Specifically, the Adam optimizer is applied with a decaying learning rate schedule: $1 \times 10^{-3}$ for the first 20,000 epochs, $5\times10^{-4}$ for the next 20,000 epochs, and $1 \times 10^{-4}$ for the final 20,000 epochs. To further refine the solution, the L-BFGS optimizer is employed for up to 50,000 additional epochs, ensuring the residuals are minimized.  
An approximate solution is obtained once the loss function reaches a sufficiently low level. To assess prediction accuracy, we use the relative $L^2$ error as a performance metric:  
\begin{equation}  
{\varepsilon _V} = \frac{{{{\left\| {V - {V^*}} \right\|}_2}}}{{{{\left\| {{V^*}} \right\|}_2}}},  
\end{equation}  
where $V$ represents the predicted values, and $V^*$ denotes the reference values.

After training, the boundary shape is extracted from the predicted body fraction field \(\phi\). Since \(\phi = 1\) represents the solid region and \(\phi = 0\) represents the fluid region, there exist intermediate values between 0 and 1. 
To define the solid region, we first identify all inference points within the target reconstruction domain where \(\phi\) lies within a carefully selected range (from a specific value up to 1). The selection of this threshold value requires fine-tuning to ensure an accurate boundary extraction. 
This process allows us to obtain all points within the solid region, including those on the boundary.
Once the solid region is determined, a convex hull algorithm~\cite{barber1996quickhull} is applied to identify the outermost layer of these points, forming the estimated boundary shape of the hidden body.

All implementations are built upon the DeepXDE framework~\cite{lu2021deepxde}, which utilizes TensorFlow~\cite{abadi2016tensorflow} as its backend.  
For the inverse problem, we incorporate high-fidelity direct numerical simulation (DNS) data generated using OpenFOAM, a finite volume method (FVM)-based solver~\cite{jasak2009openfoam}. 
These DNS results serve as partial input data for inference and are also used as reference solutions in Sec.~\ref{sec:Numerical results}.

\section{Numerical results and discussions}
\label{sec:Numerical results}
In this section, we validate the proposed PINN-based approach across three classical flow cases, validating its effectiveness in different scenarios. These cases include (i) a steady flow around a fixed cylinder, representing a stationary boundary in an incompressible flow, (ii) an in-line oscillating cylinder in fluid, introducing the challenge of inferring a moving boundary, and (iii) subsonic inviscid flow over an airfoil, highlighting the application of this approach in compressible flow. 
Numerical experiments for each case are conducted to assess the accuracy and robustness of the approach.  

\subsection{Steady flow around a fixed cylinder}
\label{subsec:steady flow around a fixed cylinder}
To validate the feasibility of the proposed approach, we first consider a simple benchmark case: a steady flow around a fixed cylinder at a Reynolds number of $Re = 50$. This case serves as a fundamental test to assess the capacity of the approach of reconstructing both the flow field and the solid boundary using sparse velocity measurements. With this Reynolds number, the flow remains steady and exhibits a symmetrical recirculation region behind the cylinder without vortex shedding. By applying our proposed approach to this scenario, we aim to evaluate whether the inferred boundary and the reconstructed flow fields are consistent with the expected physical characteristics of the flow.

In this case, the actual body is a cylinder centered at the origin. However, its shape and location are unknown in the inverse setting.
For this inverse problem, the computational domain $\Omega$ is defined as a square region \([-1.5, 1.5] \times [-1.5, 1.5]\), with the radius of the cylinder taken as the unit length. The target region, $\Omega_1$, which contains the hidden body, is defined as a smaller square domain of \([-0.65, 0.65] \times [-0.65, 0.65]\), where no information is available apart from the existence of a stationary body. The remaining area in $\Omega$ (excluding $\Omega_1$) is denoted as $\Omega_2$, where sparse velocity measurements can be sampled. The objective is to infer the shape and position of the hidden boundary using the sampled velocity data $u$ and $v$ from $\Omega_2$ and to reconstruct the complete flow field ($u, v, p$) within $\Omega$.  
For training, sparse data points are randomly selected from $\Omega_2$ at three different resolutions: 1000, 500, and 100 points. To impose the physics-informed constraints, 50,000 collocation points are randomly distributed across the entire domain $\Omega$, ensuring sufficient coverage for accurate reconstruction.
\begin{figure*}[htb!]
    \centering
    \includegraphics[width=0.9\linewidth]{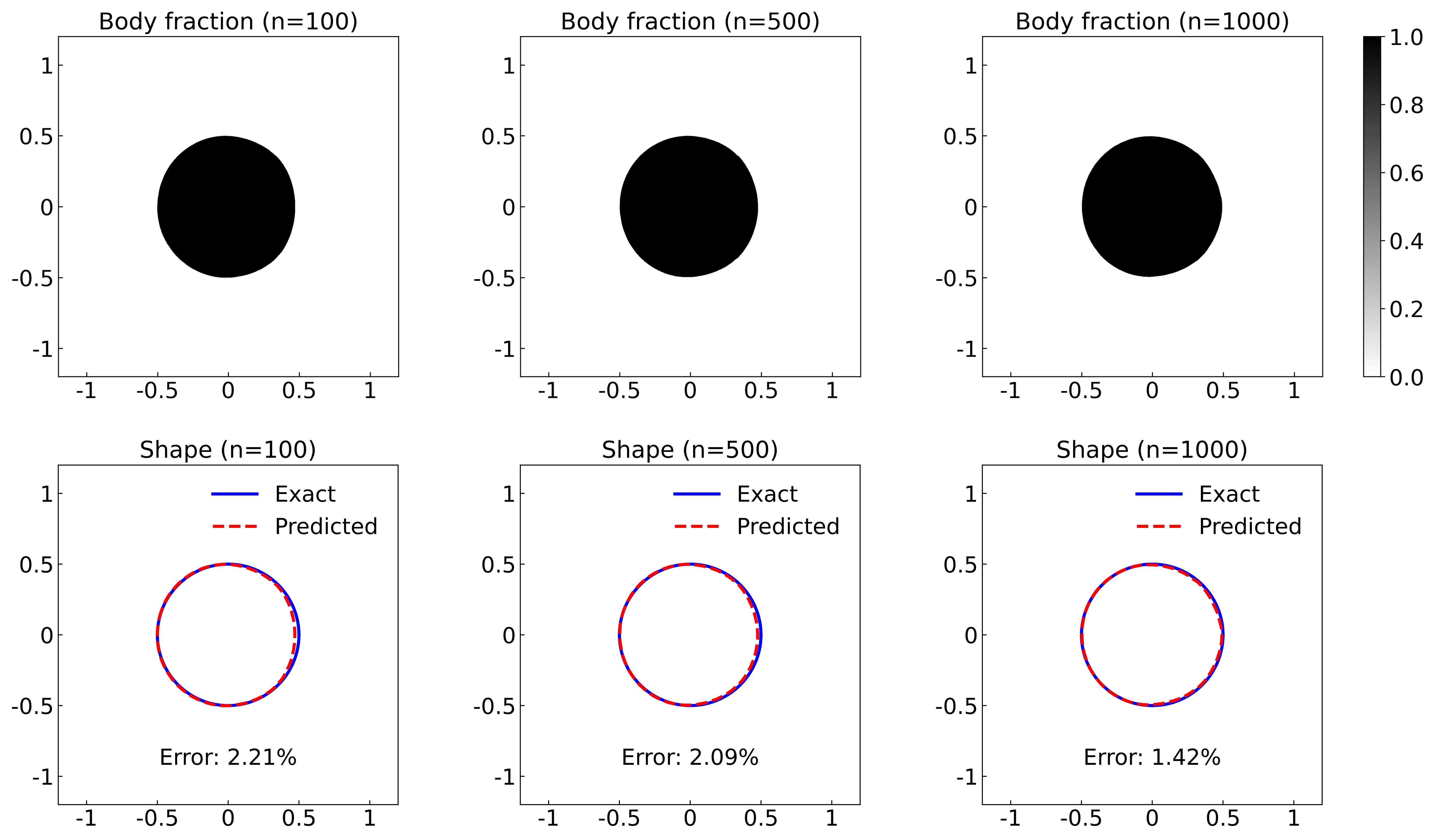}
    \caption{
    Predicted body fraction $\phi$ distributions and inferred boundary shapes for the steady flow around a fixed cylinder at three data resolutions (100, 500, and 1000 randomly sampled points). The first row shows the predicted body fraction distributions, while the second row presents the comparison between the predicted and the exact cylinder boundary shapes, along with the corresponding relative $L^2$ errors.  
    }
    \label{fig:case1_inferred_BF_shape}
\end{figure*}
\begin{figure*}[htb!]
    \centering
    \includegraphics[width=0.7\linewidth]{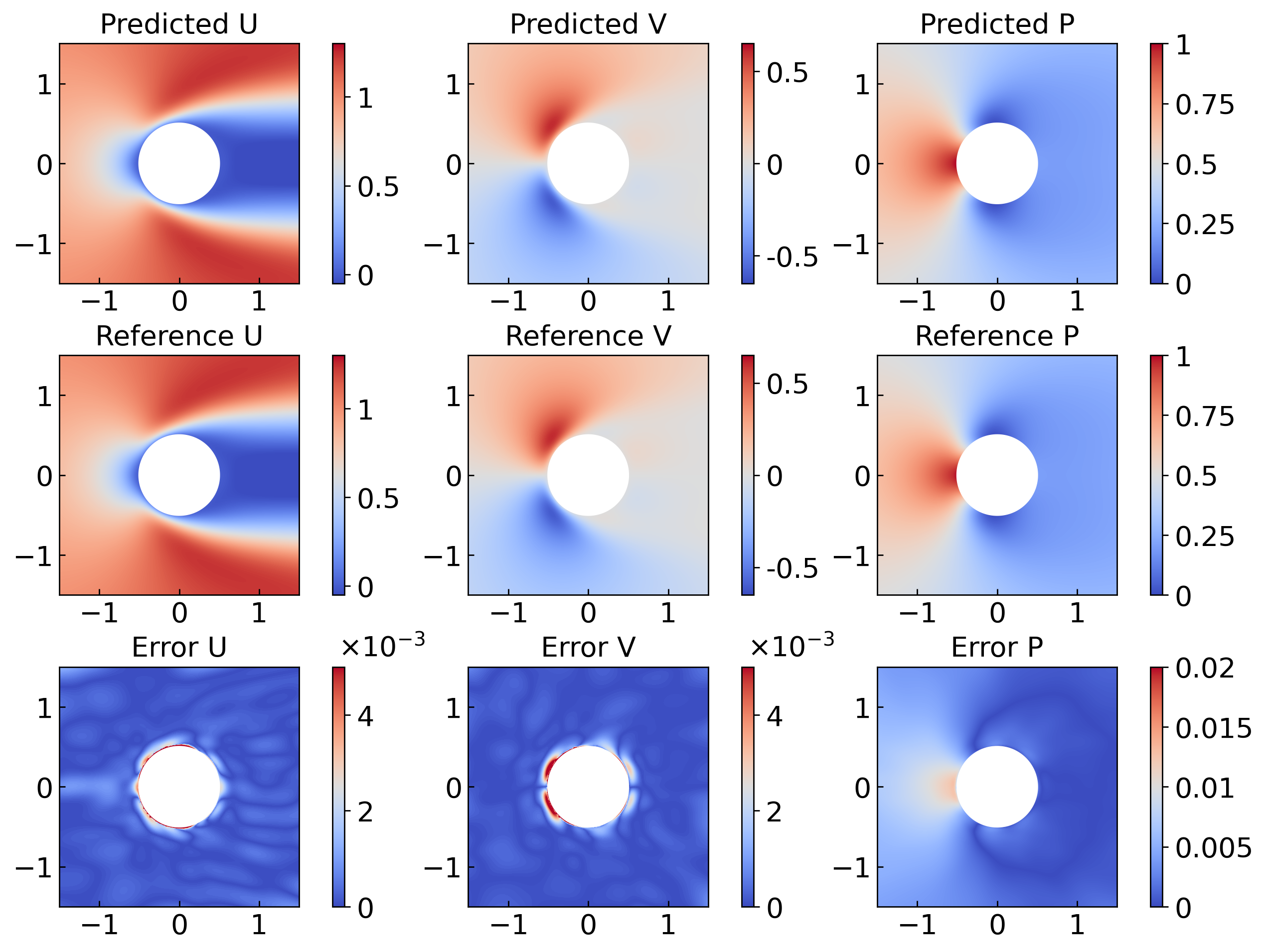}
    \caption{Comparison between the predicted~(trained with 100 random data points) and the reference flow fields for the velocity $u$, $v$, and pressure $p$ in the steady flow around a fixed cylinder. 
    The bottom row depicts the point-wise absolute errors. }
    \label{fig:case1_inferred_flows}
\end{figure*}

Fig.~\ref{fig:case1_inferred_BF_shape} displays the results of the inferred cylinder boundary under different data resolutions. 
The first row shows the predicted body fraction $\phi$ distributions for models trained with 100, 500, and 1000 randomly sampled velocity data points, where the three inferred bodies are consistently in clear shapes of cylinder. 
In this case, the specific threshold is set to 0.9~(with the range \(\phi = [0.9, 1]\)) to effectively extract the boundary shape. 
These distributions show the adjusted \(\phi\) values based on the identified boundary, where \(\phi = 1\) in the solid region and \(\phi = 0\) in the fluid region.
The second row of Fig.~\ref{fig:case1_inferred_BF_shape} compares the predicted boundary shapes with the exact cylinder boundary, along with the corresponding relative $L^2$ errors. 
The relative $L^2$ errors of the predicted cylinder shapes are computed by measuring the distance between each boundary point of the predicted shape and its nearest neighbor on the exact cylinder. This is equivalent to evaluating the discrepancy between the predicted and the actual radius distributions.
From such comparisons, it is evident that the front and sides of the cylinder are in consistent with the exact boundary, while the primary discrepancies occur at the rear. A possible explanation is that the flow velocity behind the cylinder is relatively low, which leads to a small value of the solid term $\alpha \phi \bm u$ in the governing equations. This creates a competitive relationship between the solid and fluid terms during optimization. Additionally, the sparse measurement points provide limited information about the wake region, making it more challenging to accurately reconstruct the rear boundary. Although in this case, we have set the penalty coefficient $\alpha$ to 10 to balance this effect, some minor deviations in the inferred shape still remain.  
As the number of data points increases, the prediction near the rear side of the cylinder becomes more accurate, reducing shape reconstruction errors.

Fig.~\ref{fig:case1_inferred_flows} presents a comparison between the predicted and the reference flow fields for the velocity $u$, $v$, and pressure $p$. 
And the point-wise absolute errors are also depicted in the bottom row of Fig.~\ref{fig:case1_inferred_flows}.
The point-wise absolute errors show that the predicted results are in consistent with the reference solution, with only minor discrepancies observed near the cylinder surface. 
The relative \( L^2 \) errors of the reconstructed flow field within the entire domain $\Omega$ are 0.14\% for \( u \), 0.49\% for \( v \), and 0.94\% for \( p \).
The predictions are obtained from a model trained with only 100 randomly sampled data points, demonstrating the good capability of the model to both infer the stationary boundary and reconstruct the surrounding flow field with sparse measurements. 

\subsection{In-line oscillating cylinder in fluid}
\label{subsec:In-line oscillating cylinder}
\begin{figure*}[htb!]
    \centering
    \includegraphics[width=1.0\linewidth]{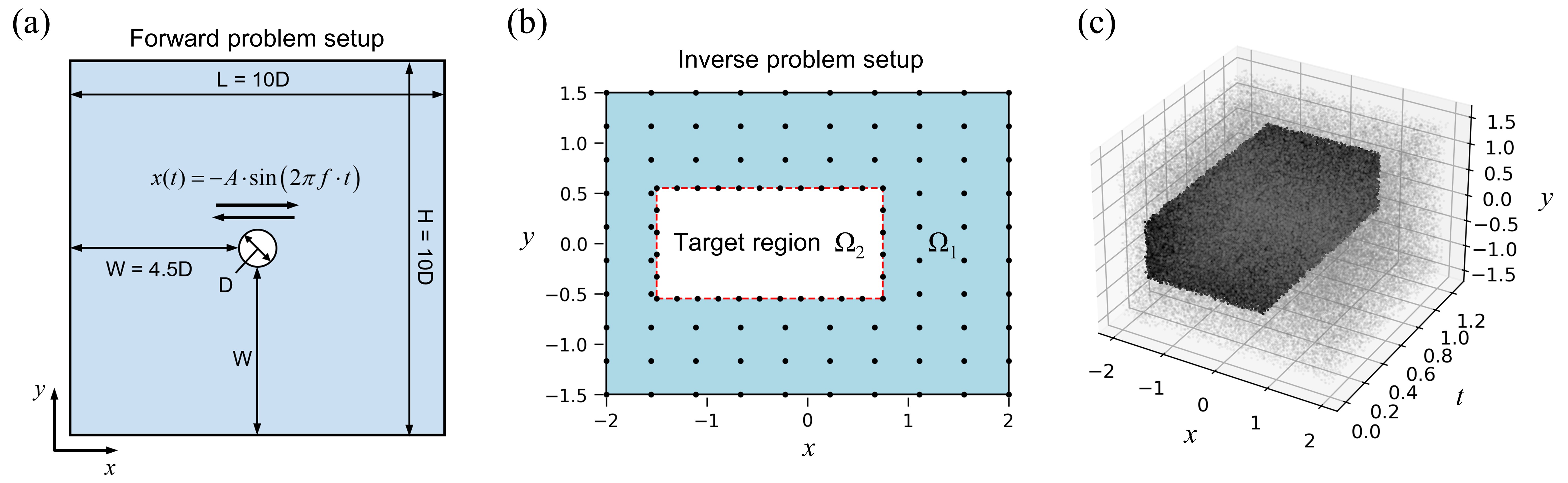}
    \caption{The forward and inverse problem setups, along with the distribution of training points, for the in-line oscillating cylinder in fluid.  
    (a) Computational domain and prescribed motion of the oscillating cylinder for the forward problem, following the configuration in Ref.~\cite{zhu2024physics}.  
    (b) Computational domain for the inverse problem (where \( \Omega \) represents the whole region, \( \Omega_1 \) denotes the region with available velocity measurements, and \( \Omega_2 \) is the target reconstruction region containing a hidden moving boundary) with spatial distribution of measurement data points~($10 \times 10$) in \( \Omega_1 \) and 30 extra sensor measurement points  uniformly placed along the boundary of \( \Omega_2 \).  
    (c) Spatial-temporal distribution of randomly sampled collocation points (in \( \Omega \) and \( \Omega_2 \)), used for enforcing physical constraints.}
    \label{fig:inline_cyl_setup}
\end{figure*}

The motion of an oscillating cylinder in a quiescent fluid is a classical scenario in fluid-structure interaction. This scenario has been extensively studied in the literature.~\cite{dutsch1998low, liu2014efficient, zhu2024physics}, as it involves complex vortex-structure interactions.  
Two key dimensionless parameters characterize the flow dynamics: the Reynolds number ($Re$) and the Keulegan-Carpenter number ($KC$), defined as:  
\begin{equation}
Re = \frac{{U_{\max} D}}{\nu},\quad KC = \frac{{U_{\max}}}{fD},
\end{equation}  
where $U_{max}$ is the peak velocity of the oscillating cylinder, $D$ is its diameter, $\nu$ is the fluid's kinematic viscosity, and $f$ is the oscillation frequency. 
The prescribed motion of the cylinder follows simple harmonic oscillation:  
\begin{equation}
x(t) = - A \sin (2\pi f \cdot t),
\end{equation}  
where $x$ represents the center position of the cylinder, and $A$ is the oscillation amplitude. The Keulegan-Carpenter number can alternatively be written as $KC = \frac{2 \pi A}{D}$.
Therefore, the reference function for the velocity of the moving cylinder is $u(t)=-2\pi f \cdot A\cdot\cos(2\pi f\cdot t)$ and $v(t)=0$.
The computational domain and the prescribed motion of the cylinder for solving the forward problem are configured according to their setup, as illustrated in Fig.~\ref{fig:inline_cyl_setup}(a).
Following the numerical simulations of Zhu et al.~\cite{zhu2024physics}, we respectively set the Reynolds and Keulegan-Carpenter numbers to $Re=125$ and $KC=5$. The corresponding physical parameters are chosen as $\rho=1$, $\nu=0.008$, $D=1$, and ${U_{max}}=1$, with an oscillation period of $T=1/f=5$. 

For the inverse problem, the computational domain is illustrated in Fig.~\ref{fig:inline_cyl_setup}(b). The entire flow field region $[-2, 2] \times [-1.5, 1.5]$ is denoted as $\Omega$, with the radius of the cylinder taken as the unit length. Within this domain, $\Omega_1$ represents the region where velocity sensors are placed, providing sparse measurements of velocity components $(u, v)$ at different time frames. The target reconstruction region, $\Omega_2$ $[-1.5, 0.75] \times [-0.55, 0.55]$ contains an unknown flow field where a hidden moving boundary exists, satisfying the no-slip condition. The objective of the inverse problem is to infer the velocity and trajectory of this hidden boundary using the sparse velocity measurements from $\Omega_1$, while also reconstructing the complete flow field, including velocity $(u, v)$ and pressure $p$, within $\Omega_1$ and $\Omega_2$. 
The time domain for the inverse problem is chosen as one-quarter of the oscillation period $T$ of the moving cylinder, i.e., [0, 1.25].

The training points for the inverse problem are configured as follows. 
The data points, representing sensor measurements, are uniformly sampled within the \( \Omega_1 \) region at three different spatial resolutions: \( 30 \times 30 \), \( 20 \times 20 \), and \( 10 \times 10 \). The sampling process begins with a uniform distribution of sensors across the entire domain \( \Omega \) at the specified resolution. However, any points falling within the target reconstruction region \( \Omega_2 \) are removed, ensuring that only measurement points within \( \Omega_1 \) are retained for training. 
Apart from the data points sampled in the above region, 30 extra sensor measurement points are uniformly placed along the boundary of \( \Omega_2 \), providing additional information at the interface between the known and unknown regions. This additional boundary sampling is applied consistently across all three resolution levels.
Velocity data are sampled every $\Delta t =0.05$, resulting in a total of 26 time frames used for training.
In addition to data points, collocation points are used to enforce the physical constraints. To achieve sufficient coverage, we randomly sample 100,000 collocation points within the spatial-temporal domain \( \Omega \) and an additional 150,000 points specifically within the target spatial-temporal domain \( \Omega_2 \). This ensures that the model learns both from the known flow regions and the hidden boundary region where inference is required. Fig.~\ref{fig:inline_cyl_setup}(b) and (c) illustrate the spatial distribution of the measurement data points~($10 \times 10$) and the sampled collocation points.

\begin{figure*}[htb!]
    \centering
    \includegraphics[width=1.0\linewidth]{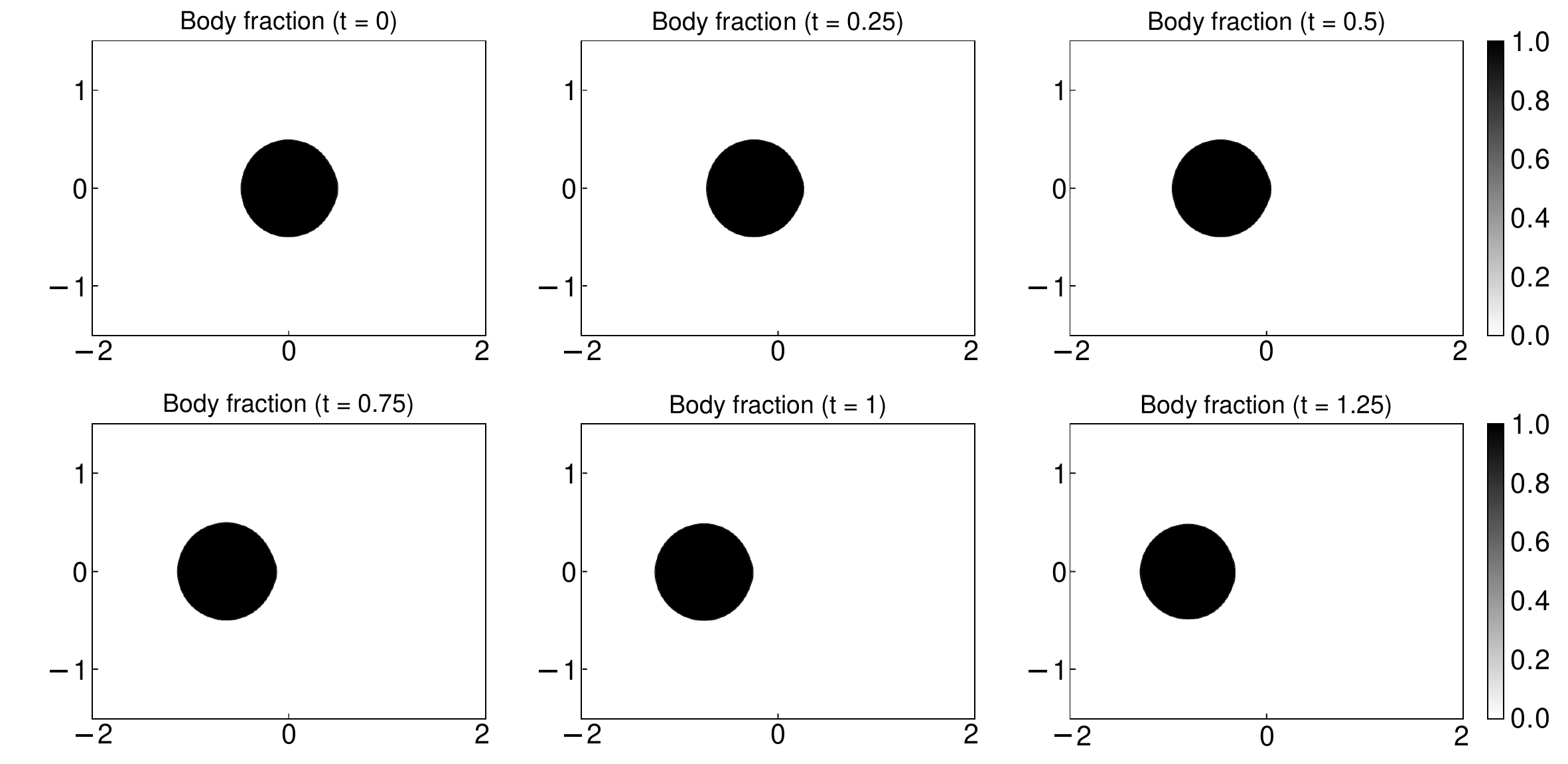}
    \caption{Time evolution of the predicted body fraction $\phi$ distribution at different time instances~($\Delta t=0.25$). The results are obtained from a model trained with velocity measurements sampled at a \( 20 \times 20 \) resolution. The inferred body exhibits a clear cylindrical boundary at each time step and moves consistently with the prescribed trajectory.}
    \label{fig:case2_inferred_BF}
\end{figure*}

\begin{figure*}[htb!]
    \centering
    \includegraphics[width=1.0\linewidth]{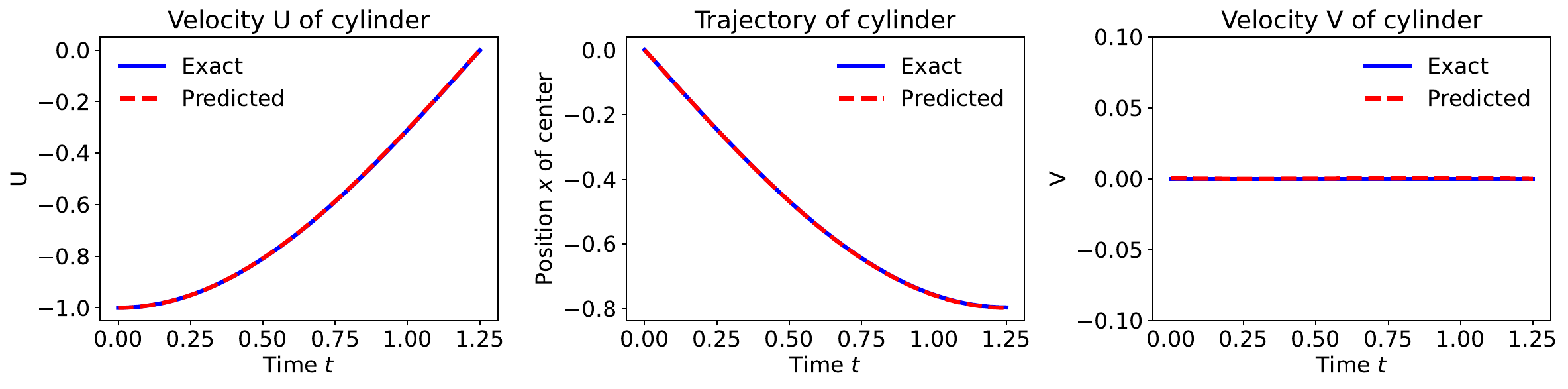}
    \caption{Predicted motion of the oscillating cylinder inferred using PINNs trained with velocity measurements at a \( 20 \times 20 \) resolution. The left, middle, and right subplots show (left) the evolution of velocity \( U \), (middle) the trajectory of the cylinder's $x$-position, and (right) the evolution of velocity \( V \), with both the predicted and the exact values. The results indicate that \( U \) aligns well with the prescribed motion, while $V$ remains near zero, confirming that the cylinder moves only in the $x$-direction.}
    \label{fig:case2_inferred_cyl_velocity}
\end{figure*}

Fig.~\ref{fig:case2_inferred_BF} shows the predicted body fraction distributions at six different time instances~($\Delta t=0.25$), providing the evolution of location of the inferred moving boundary. These results are obtained from a model trained using velocity measurements sampled at a \( 20 \times 20 \) resolution.
These distributions show the adjusted \(\phi\) values based on the identified boundary.
The body fraction distributions have a sharp distinction between the solid and fluid regions. The interior of the boundary appears black, indicating a value of 1 for the solid, while the exterior is white, representing a value of 0 for the fluid.
The inferred body exhibits a clear cylindrical boundary at each time step and moves consistently with the prescribed trajectory. 
We present the inferred motion of the oscillating cylinder using the PINN model trained on velocity measurements sampled at a \( 20 \times 20 \) resolution in Fig.~\ref{fig:case2_inferred_cyl_velocity}. It consists of three subplots: the evolution of the cylinder's velocity \( U \) over time, the trajectory of the cylinder's $x$-position, and the evolution of velocity \( V \) over time. Each plot compares the predicted results with the exact values.  
The inferred velocity \( U \) is consistent with the prescribed motion from the forward problem, while \( V \) remains consistently at or near zero, confirming that the cylinder moves strictly along the $x$-direction with no displacement in $y$. The cylinder's trajectory in the $x$-direction can be determined by first inferring its initial position from the reconstructed boundary shape and then integrating the velocity \( U \) over time. 
The predicted motion exhibits excellent agreement with the exact motion, demonstrating the good capability of the model to accurately infer the motion of the hidden moving boundary.
\begin{figure*}[htb!]
    \centering
    \includegraphics[width=0.9\linewidth]{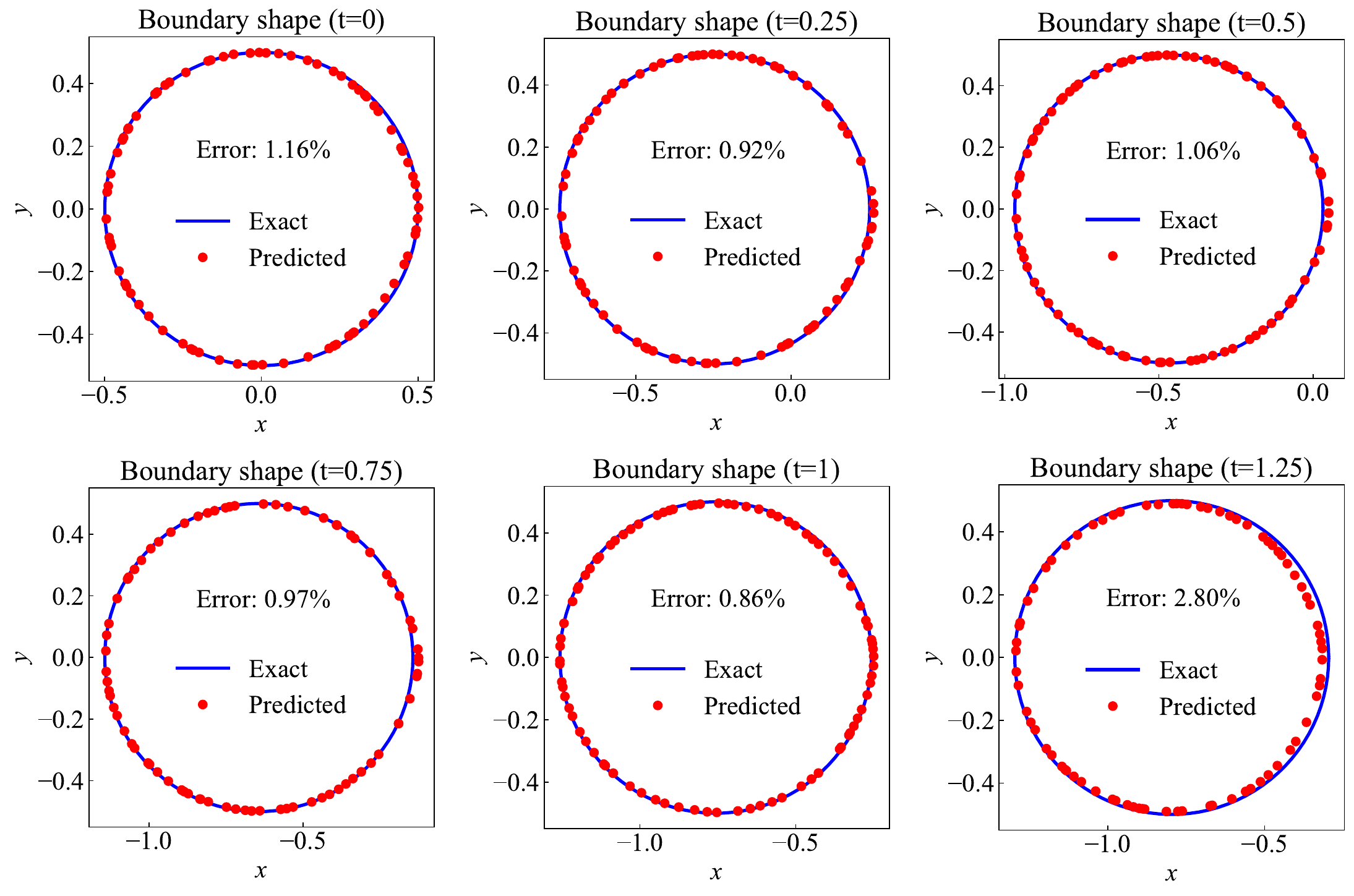}
    \caption{Comparison between the inferred and actual boundary shapes of the moving cylinder at six different time instances, along with the corresponding relative $L^2$ error.}
    \label{fig:case2_inferred_shape}
\end{figure*}
\begin{figure*}[htb!]
    \centering
    \includegraphics[width=1.0\linewidth]{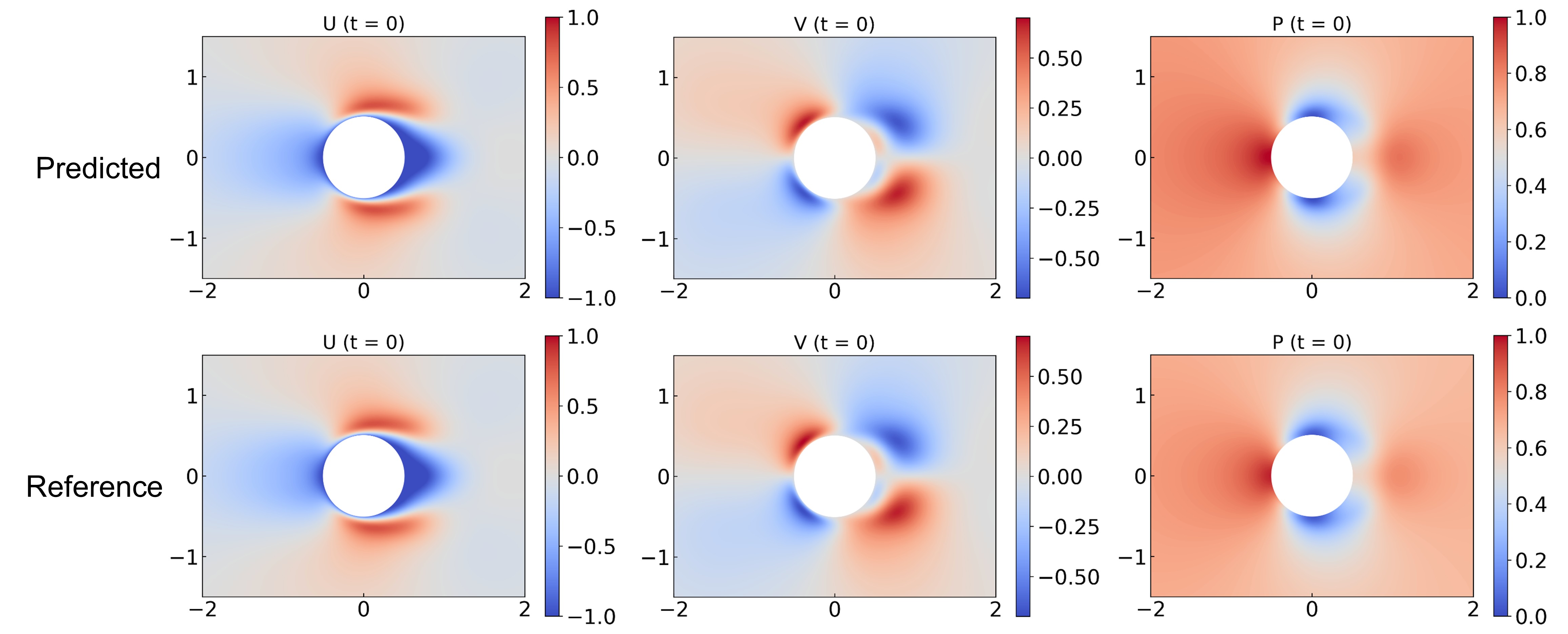}
    \includegraphics[width=1.0\linewidth]{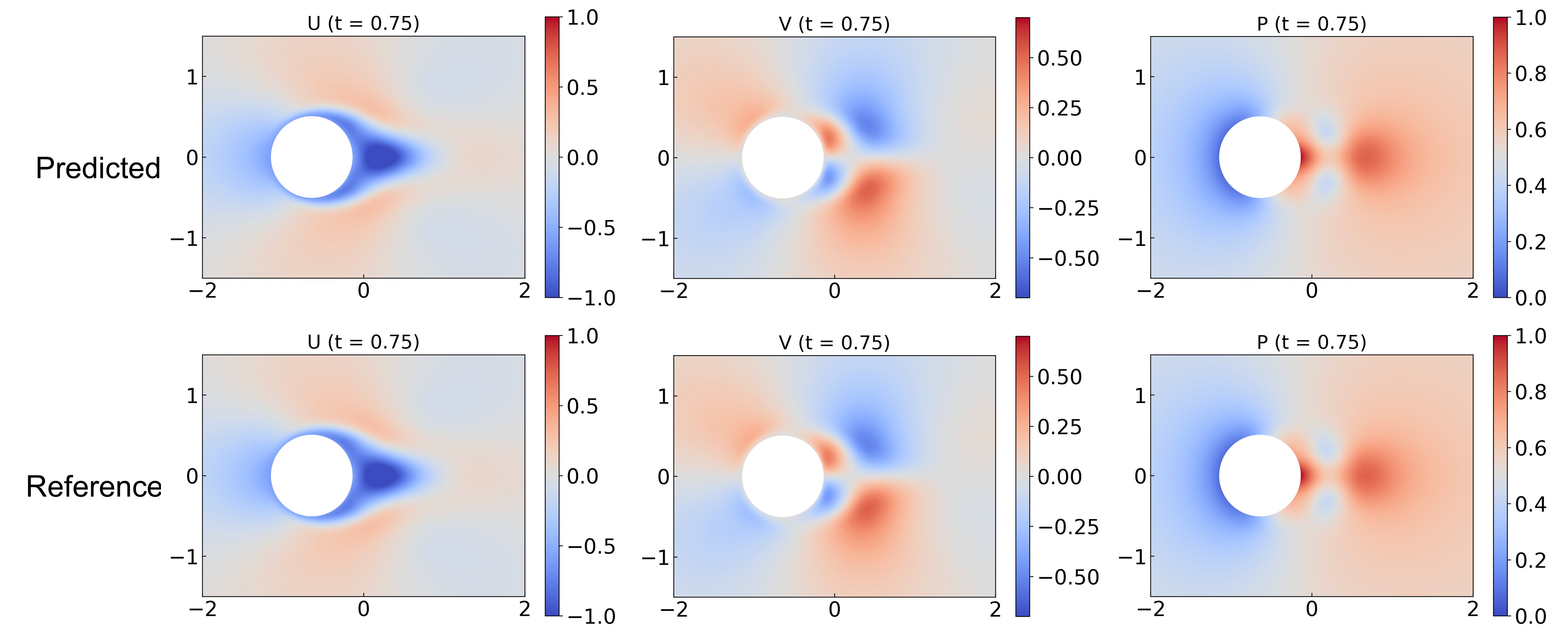}
    \caption{Comparison between the predicted and reference flow fields for the velocity $u$, $v$ and pressure $p$ at $t$=0 and $t$=0.75 for the in-line moving cylinder in fluid.}
    \label{fig:case2_inferred_flows}
\end{figure*}

Fig.~\ref{fig:case2_inferred_shape} compares the inferred boundary shape of the moving cylinder, obtained from the PINN model trained with velocity measurements at a \( 20 \times 20 \) resolution, against the actual boundary shape at six different time instances~($\Delta t=0.25$). These inferred shapes correspond to the body fraction distributions shown in Fig.~\ref{fig:case2_inferred_BF}, processed to extract the boundary contours. 
In this case, the specific threshold is set to 0.99~(with the range \(\phi = [0.99, 1]\)) to effectively extract the boundary shape.
Additionally, the relative $L^2$ error between the predicted and actual boundary shapes is presented.
The relative $L^2$ errors of the predicted cylinder shapes are computed by measuring the distance between each boundary point of the predicted shape and its nearest neighbor on the exact cylinder. 
The results demonstrate that the inferred boundary closely match the actual cylindrical shape at all time steps.  

\begin{figure*}[htb!]
    \centering
    \includegraphics[width=0.9\linewidth]{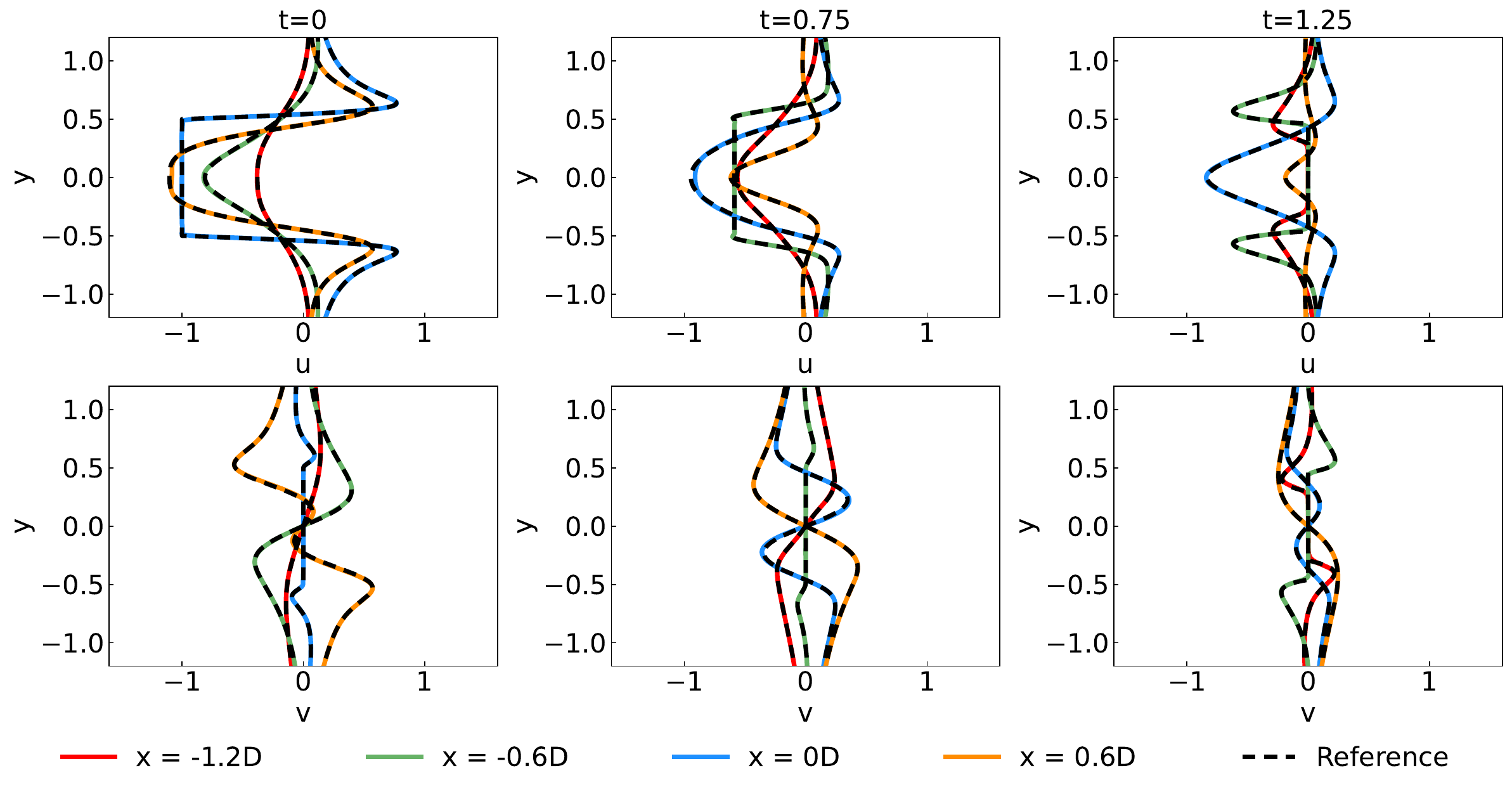}
    \caption{Comparison of the predicted and reference velocity profiles for $u$ and $v$ along the $y$-direction at three time distances~($t$=0, 0.75 and 1.25) for the in-line moving cylinder. Solid and dotted lines denote predicted and reference results, respectively, along four vertical profiles at different streamwise locations.
    }
    \label{fig:case2_uv_profiles}
\end{figure*}

Fig.~\ref{fig:case2_inferred_flows} shows a comparison between the predicted and reference flow fields for velocity components $u$, $v$ and pressure $p$ at two different time instances, $t$=0 and $t$=0.75. 
The predicted results are obtained using PINNs trained with $20 \times 20$ resolution of the velocity measurements, while the reference fields are taken from the forward simulation. 
The reconstructed flow field in $\Omega$ captures the key flow features and exhibits excellent consistency with the reference solution. The velocity contours show smooth variations and correctly depict the interaction between the moving boundary and the surrounding fluid.
To further evaluate the accuracy of the reconstructed flow field, the velocity profiles for $u$ and $v$ along the $y$-direction at three time instances ($t$=0, 0.75 and 1.25) are shown in Fig.~\ref{fig:case2_uv_profiles}. The predicted results are compared with the reference data along four vertical cross-sections. 
The comparison shows that the proposed model accurately captures the velocity distribution across different locations and time steps, demonstrating its ability to reconstruct fine flow structures with high precision.  
\begin{figure*}[htb!]
    \centering
    \includegraphics[width=1.0\linewidth]{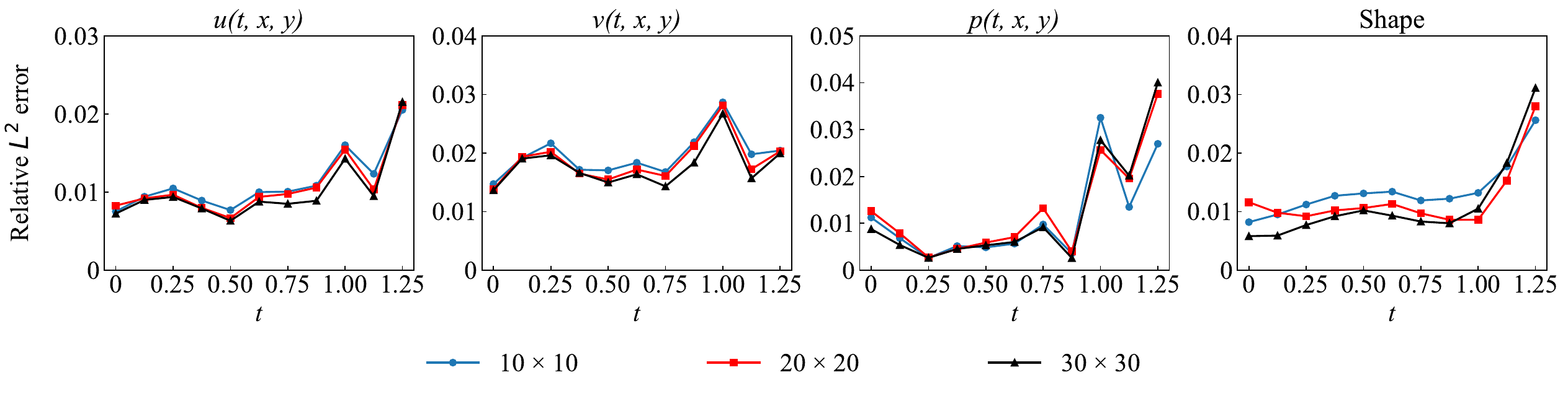}
    \caption{Relative $L^2$ errors of the predicted velocity, pressure, and inferred boundary shape for models trained with velocity measurements at three different spatial resolutions.}
    \label{fig:case2_inferred_uvps_l2_error}
\end{figure*}

We provide the relative $L^2$ errors of the predicted velocity, pressure, and inferred boundary shape obtained from models trained with velocity measurements at three different spatial resolutions in Fig.~\ref{fig:case2_inferred_uvps_l2_error}. 
The uniformly sampled points in the target reconstruction region $\Omega_2$ are used as the coordinate points for the prediction and validation of the models.
The results reveal that the velocity predictions exhibit relatively low errors, with both $u$ and $v$ maintaining an error below 3\% across all time instances, with $v$ generally exhibiting slightly higher errors than $u$.
Moreover, the velocity reconstruction is relatively insensitive to the measurement resolution, as even the sparsest $10 \times 10$ dataset achieves high accuracy in capturing the flow field. This may be due to the additional constraints imposed by velocity measurements on the velocity terms, which enhance the reconstruction of the velocity field.
In contrast, the pressure predictions exhibit slightly lower errors compared to velocity before $t = 1$, but show a noticeable increase afterward.
This may be attributed to the deceleration of the cylinder after $t = 1$, which induces rapid pressure variations in its wake. As a result, the pressure field during this period requires finer data sampling to maintain accuracy, leading to a more pronounced error increase in the coarser datasets.  
The accuracy of the inferred boundary shape remains low sensitive to the resolution of the velocity data. While the model trained with the sparsest $10 \times 10$ dataset shows a slight increase in reconstruction error, it still achieves reasonably accurate results, particularly during the time period before $t = 1$.
Additionally, the shape inference error follows a similar temporal trend across different resolutions, indicating that the overall reconstruction quality is maintained regardless of measurement density. This suggests that when data is limited or sparsely distributed, the physical constraints play a critical compensatory role in ensuring accurate reconstruction.
\begin{figure*}[htb!]
    \centering
    \includegraphics[width=1.0\linewidth]{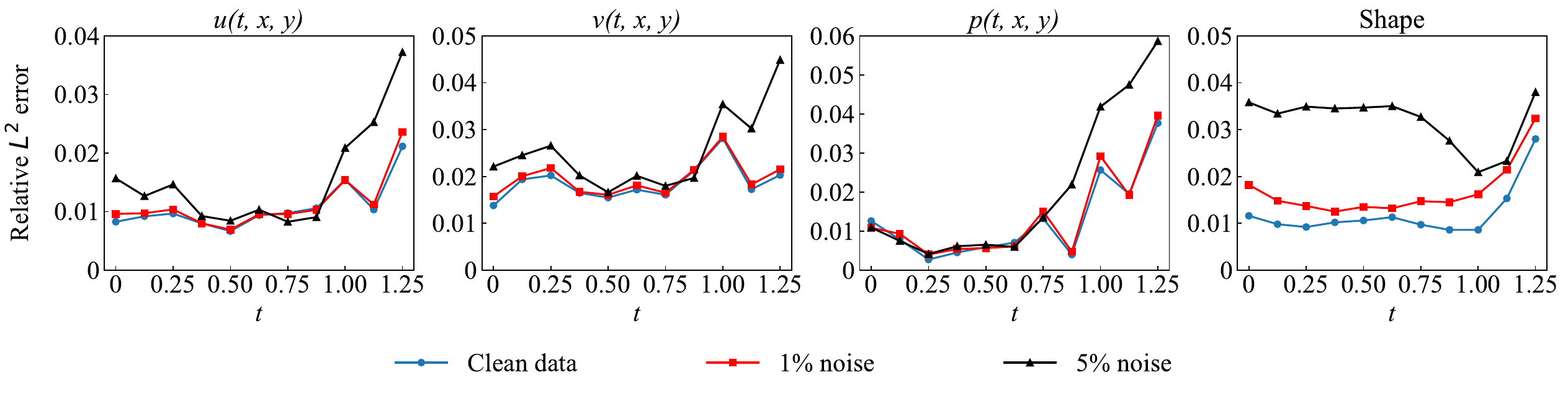}
    \caption{Relative $L^2$ errors of the predicted velocity, pressure, and inferred boundary shape for models trained with clean velocity measurements at a $20 \times 20$ resolution and data with added 1\% and 5\% noise.}
    \label{fig:case2_add_noise_l2_error}
\end{figure*}

We finally analyze the impact of data noise on the accuracy of the model in reconstructing the flow field and inferring the boundary.
Fig.~\ref{fig:case2_add_noise_l2_error} provides the relative $L^2$ errors of the predicted velocity, pressure, and inferred boundary shape under three conditions: clean velocity measurements at a $20 \times 20$ resolution, and data with added 1\% and 5\% noise. The results show that the velocity field is relatively robust to noise in the time interval $t = 0.375$ to 0.875, where the errors remain comparable across all cases. However, at both earlier and later time periods, the impact of noise becomes more pronounced, leading to a significant increase in error compared to the clean data case. Despite this, even with 5\% noise, the reconstruction of $u$ and $v$ maintains an error below 5\% at all time instances, indicating a reasonable level of accuracy.  
Similarly, the pressure field is partial sensitive to noise, particularly after $t = 0.75$, where the introduction of 5\% noise results in a substantial error increase. 
The inferred boundary shape is notably affected by noise, with both 1\% and 5\% noise levels negatively impacting the accuracy of shape reconstruction. Additionally, the shape inference error exhibits a positive correlation with noise intensity, showing that it is more sensitive to measurement noise but still maintains a reasonable level of robustness, with errors remaining below 4\% even at 5\% noise.

\subsection{Subsonic inviscid flow over an airfoil}
\label{subsec:Subsonic inviscid flow over an airfoil}
Here, we extend the proposed approach to compressible flows by solving the Euler equations and investigating an inverse problem involving subsonic inviscid flow over a NACA 0012 airfoil. The free-stream Mach number is set to $M$=0.5 with an angle of attack of $\alpha = 1.25^\circ$, ensuring that the flow remains entirely subsonic, with the Mach number consistently below one throughout the domain. Since the flow is inviscid, it exhibits smooth variations without shocks.  
We first introduce the setup of the forward simulation, whose results serve as both the data source for the inverse problem and the reference for evaluation.
The chord length of the airfoil is taken as the characteristic length scale, and the far-field boundary is positioned approximately 25 chord lengths away from the airfoil center. Following the numerical simulations of Vassberg and Jameson~\cite{vassberg2010pursuit}, we define the NACA 0012 airfoil geometry using its analytical equation~\cite{abbott2012theory}, ensuring a sharp trailing edge. At the far-field boundary, a steady free-stream condition with $M$=0.5 and $\alpha = 1.25^\circ$ is imposed as a Dirichlet boundary condition, while slip boundary conditions are applied along the airfoil surface.  

\begin{figure*}[htb]
    \centering
    \includegraphics[width=0.8\linewidth]{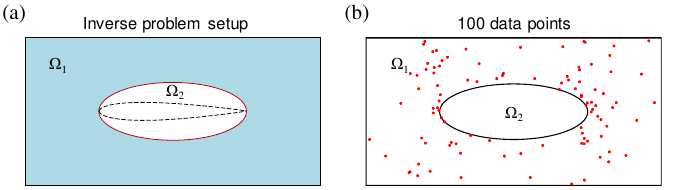}
    \caption{Setup of the inverse problem and distribution of training points for the subsonic inviscid flow over a NACA 0012 airfoil.
    (a) Geometric domain, where $\Omega$ represents the whole region, $\Omega_1$ denotes the region with available data points, and $\Omega_2$ is the target reconstruction region containing a hidden airfoil boundary.  
    (b) Spatial distribution of 100 randomly sampled data points from the grid nodes of the forward simulation within $\Omega_1$.}
    \label{fig:case3_setup}
\end{figure*}

For the inverse problem, the geometric domain is shown in Fig.~\ref{fig:case3_setup}(a). The entire flow field region$[-0.5, 1.5] \times [-0.5, 0.5]$ is denoted as $\Omega$, with the chord length of the airfoil taken as the unit length. Within this domain, $\Omega_1$ represents the region where sparse data points containing flow information at different locations are available. The target reconstruction region $\Omega_2$ is an elliptical domain with a major axis of 1 and a minor axis of 0.375, where the flow field is unknown and a hidden boundary satisfies the no-penetration condition. The objective of the inverse problem is to infer the boundary shape of NACA 0012 airfoil using sparse flow data from $\Omega_1$, while also reconstructing the complete flow field, including velocity $(u, v)$, pressure $p$, and density $\rho$, within both $\Omega_1$ and $\Omega_2$.  
The training points for the inverse problem are configured as follows. Available data points, randomly sampled from the grid nodes of the forward FVM simulation within $\Omega_1$, are considered at four different spatial resolutions: 1000, 700, 400, and 100 points. Additionally, collocation points are used to enforce the physics-informed constraints, with 40,000 points randomly sampled within the full domain $\Omega$ to ensure adequate coverage. Fig.~\ref{fig:case3_setup}(b) illustrates the spatial distribution of the 100 sampled data points for clarity.

\begin{figure*}[htb!]
    \centering
    \includegraphics[width=0.9\linewidth]{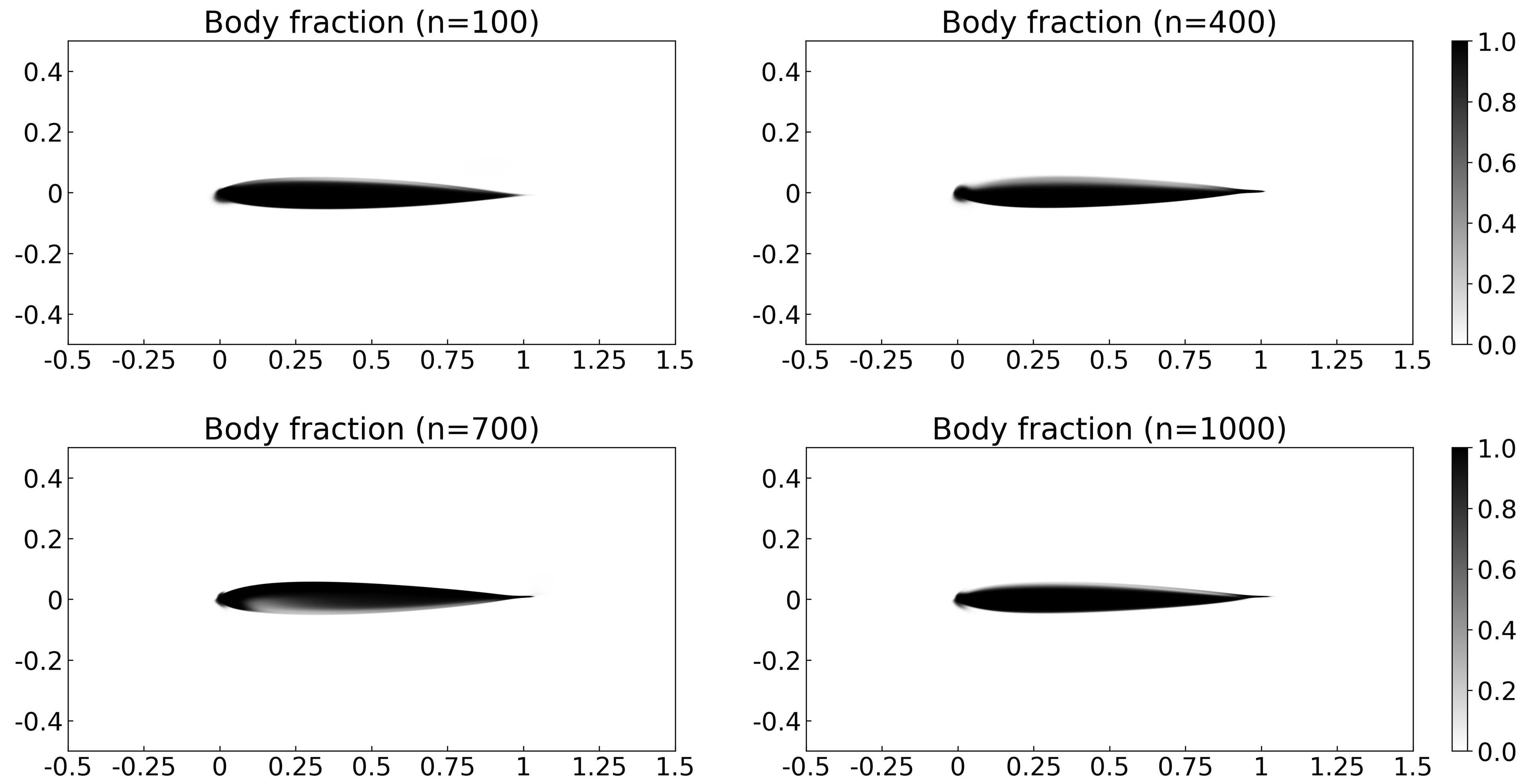}
    \caption{The predicted body fraction $\phi$ distributions for the subsonic inviscid flow over an airfoil at four different data resolutions, corresponding to 100, 400, 700, and 1000 randomly sampled flow field points. The inferred obstacle consistently exhibits a clear airfoil-shaped boundary across all cases, with minor uncertainties near the leading-edge upper or lower surfaces.}
    \label{fig:case3_inferred_BF}
\end{figure*}

Fig.~\ref{fig:case3_inferred_BF} presents the predicted body fraction $\phi$ distributions at four different resolutions of randomly sampled flow field data, specifically 100, 400, 700, and 1000 data points. 
Using the finer possible inference points for prediction within the $\Omega_2$ region can lead to a more accurate and rational distribution of the body fraction.
Across all cases, the inferred obstacle successfully reconstructs a clear airfoil-shaped boundary. While slight uncertainties are observed near the upper and lower edges of the leading section, the overall contour remains clearly identifiable. 

In this case, the specific threshold is set to 0.1~(with the range \(\phi = [0.1, 1]\)) to effectively extract the boundary shape.
To quantitatively assess the discrepancy between the inferred airfoil shape and the exact NACA 0012 airfoil, we compute the relative $L^2$ error between their boundary point distributions, incorporating arc-length reparameterization. 
Since the two sets of points may have different sampling distributions, we first reparameterize both the predicted and the exact airfoil boundary curves using their cumulative arc length and then uniformly resample them. This ensures a one-to-one correspondence between points on the two shapes. The relative $L^2$ error is then computed based on the corresponding resampled points. 

\begin{figure*}[htb!]
    \centering
    \includegraphics[width=0.9\linewidth]{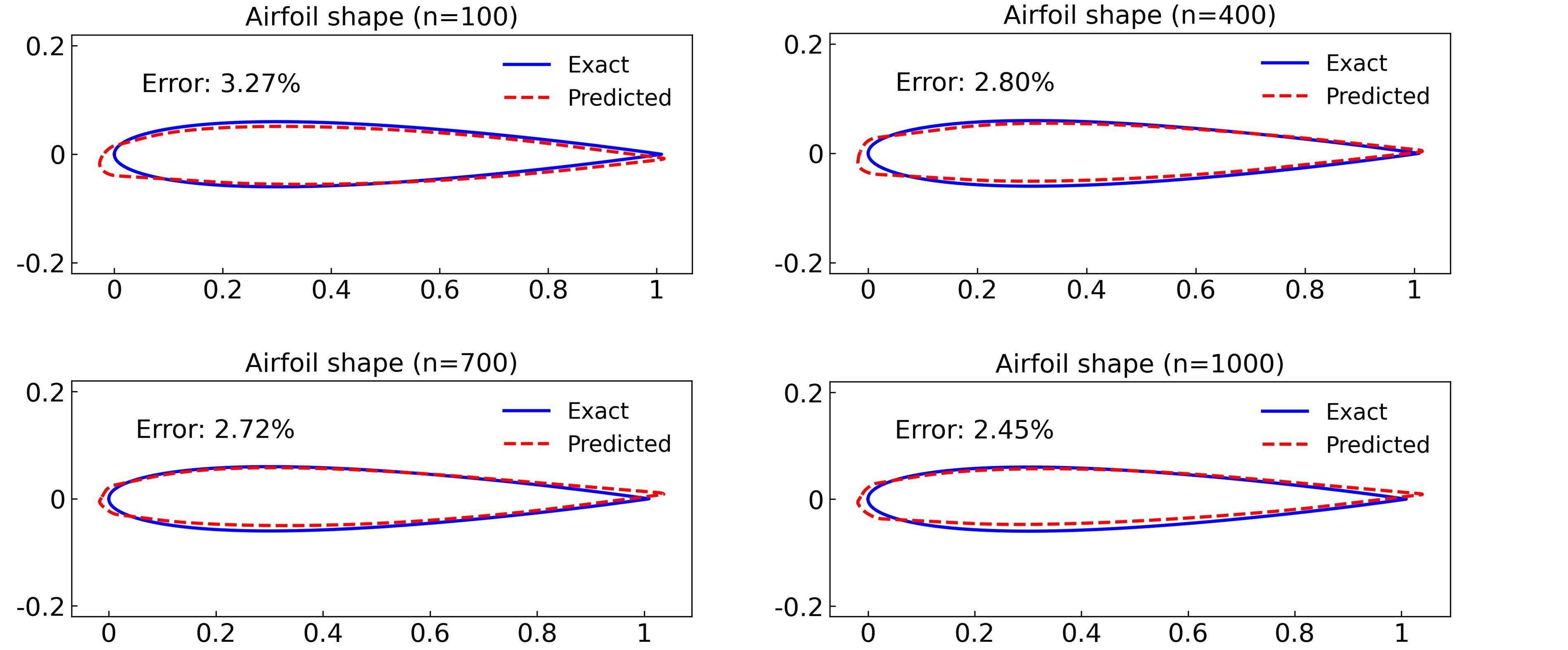}
    \caption{
    Comparison between the predicted and actual boundary shapes of NACA 0012 airfoil for models trained with 100, 400, 700, and 1000 randomly sampled data points, along with the corresponding relative $L^2$ error.}
    \label{fig:case3_inferred_shape}
\end{figure*}

\begin{figure*}[htb!]
    \centering
    \includegraphics[width=1.0\linewidth]{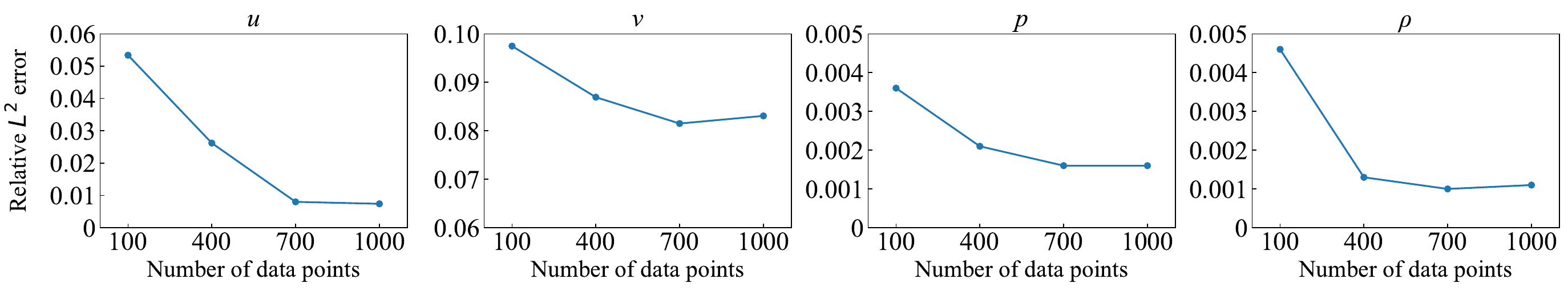}
    \caption{Relative $L^2$ errors of the predicted velocity $(u, v)$, pressure $p$, and density $\rho$ for models trained with data points at four spatial resolutions in the subsonic inviscid flow over an airfoil.}
    \label{fig:case3_inferred_uvps_l2_error}
\end{figure*}

We then provide a comparison between the predicted and actual boundary shapes of the NACA 0012 airfoil for models trained with 100, 400, 700, and 1000 randomly sampled data points in Fig.~\ref{fig:case3_inferred_shape}. Alongside the boundary shape comparison, the relative $L^2$ error between the predicted and actual boundary shapes is also presented. 
The results demonstrate that even with only 100 data points, the model is capable of reconstructing a reasonable and smooth airfoil boundary shape. 
Although there is a relatively larger deviation near the leading edge, the overall shape remains well-aligned with the exact NACA 0012 airfoil, with an error of approximately 3\%. As the number of data points increases, the discrepancy at the leading edge gradually decreases, and the overall boundary inference error is reduced. 
However, the improvement becomes less significant at higher resolutions. For instance, even with 1000 data points, the error is only reduced to 2.45\%. 
This suggests that beyond a certain data density, additional measurements provide diminishing returns in accuracy.
The diminishing improvement in boundary shape accuracy with increasing data points can be attributed to the spatial distribution of the additional samples. As more data points are introduced, they primarily populate regions at a certain distance from the airfoil rather than directly on the boundary. These new data points first enhance the accuracy of the local flow field in their immediate vicinity, and this improvement propagates toward the reconstruction of the airfoil boundary. However, the influence of these additional data points weakens as the distance increases, leading to a less pronounced enhancement in boundary shape accuracy.  
Nevertheless, it can be anticipated that the overall accuracy of the reconstructed flow field benefits more significantly from the increased data resolution. This is corroborated by Fig.~\ref{fig:case3_inferred_uvps_l2_error}, which shows a notable reduction in flow field reconstruction error as the number of data points increases.
 
\begin{figure*}[htb!]
    \centering
    \includegraphics[width=1.0\linewidth]{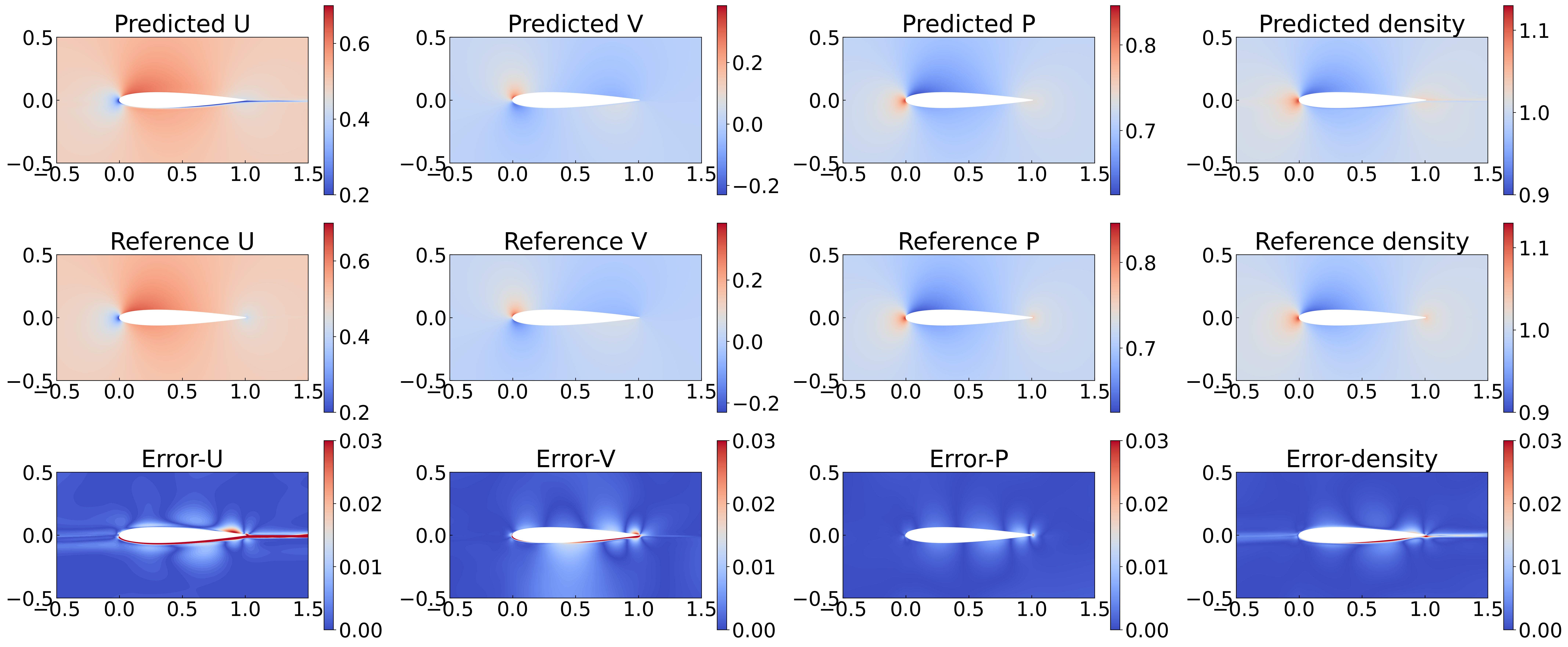}
    \caption{Comparison between the predicted~(trained with 100 random data points) and reference flow fields for the dimensionless velocity $u$, $v$, pressure $p$, and density $\rho$ in the subsonic inviscid flow over an airfoil. 
    The bottom row depicts the point-wise absolute errors. 
    The predictions are evaluated at the grid nodes of the FVM simulation.}
    \label{fig:case3_inferred_flows}
\end{figure*}

Fig.~\ref{fig:case3_inferred_flows} presents a comparison between the predicted and reference flow fields for the dimensionless velocity components $u$, $v$, pressure $p$, and density $\rho$ in the subsonic inviscid flow over an airfoil. 
And the point-wise absolute errors are also depicted in the bottom row of Fig.~\ref{fig:case3_inferred_flows}.
The predictions are evaluated at the grid nodes of the FVM simulation.
The point-wise absolute errors show that the predicted results are in consistent with the reference solution, with only minor discrepancies observed near the airfoil surface and tail. 
The predictions are obtained from a model trained with only 100 randomly sampled data points, demonstrating the capability of the model to infer both the airfoil boundary and reconstruct the surrounding flow field with sparse measurements. 
In Fig.~\ref{fig:case3_inferred_uvps_l2_error}, we provide the relative $L^2$ errors of the predicted velocity $(u, v)$, pressure $p$, and density $\rho$ for models trained with 100, 400, 700, and 1000 randomly sampled data points. 
The predictions are evaluated at the grid nodes of the FVM simulation.
The relative $L^2$ errors of the reconstructed flow field significantly decrease as the number of training data points increases. Specifically, the errors for pressure and density remain relatively low, with predictions from just 100 data points achieving errors of less than 0.5\%. The error in velocity is relatively large. For $u$, the error is highest at about 5.2\% with 100 data points, but decreases to 1\% when 1000 data points are used. For $v$, the overall error is higher compared to the other quantities, but it also shows a decrease with more data, with the maximum error not exceeding 10\%.

These results highlight the effectiveness of the proposed approach in capturing aerodynamic flow features even with sparse data conditions. More importantly, this demonstrates that the proposed method is also applicable to inferring slip no-penetration boundaries in high speed compressible flows. This exciting result suggests that the proposed approach holds potential for further exploration in other domains or for different types of boundary conditions.

\section{Conclusions}
\label{sec:Conclusions}
We have developed a novel physics-informed neural network~(PINN)-based framework that integrates a body fraction parameter into the governing equations of fluid dynamics. 
The key innovation of the approach lies in its unified treatment of fluid and solid regions: while enforcing conservation laws in fluid domains, it rigorously imposes no-slip/no-penetration conditions at solid boundaries through the body fraction formulation.
This dual enforcement enables accurate recovery of complete flow fields (velocity/pressure) and meanwhile precisely identifying boundary geometries
from sparse measurements.

We rigorously validated our framework through three canonical test cases: steady incompressible flow past a fixed cylinder, unsteady flow induced by an oscillating cylinder, and subsonic compressible flow over an airfoil.
In all scenarios, the method demonstrated consistent accuracy in reconstructing both the flow fields and boundary geometries. 
Notably, for moving boundary problems, our approach successfully captured the temporal evolution of boundary shape while precisely recovering kinematic quantities (velocity and trajectory)- a capability that surpasses conventional reconstruction methods.

We conducted systematic evaluations of our framework under challenging real-world conditions, including: sparse data, velocity-only measurements, and significant measurement noise. These tests demonstrate remarkable robustness of the method, showing that even with severely limited or noisy input data, the physics-constrained approach maintains faithful adherence to fundamental conservation laws while producing accurate reconstructions of both flow fields and embedded boundaries.

To summarize, our work makes two key contributions to flow reconstruction: (1) a novel PINN-based framework for simultaneous inference of (both stationary and moving) solid boundaries and flow fields, (2) robust performance under sparse, noisy, and velocity-only measurements. By synergistically combining sparse data with fundamental physical principles, this approach provides unprecedented insights into fluid-structure interactions while overcoming limitations of traditional methods such as data assimilation methods, where obtaining comprehensive flow measurements is often infeasible.

The proposed methodology establishes a foundation for several promising research directions. Future work could extend to three-dimensional flows, complex geometries, and multi-physics systems with additional measurement modalities (e.g., temperature or density fields). Such extensions could significantly advance real-time flow diagnostics and control capabilities in fluid dynamics.
The framework also offers transformative potential for applications ranging from aerodynamic/hydrodynamic design to biomedical flow analysis.


\section*{Conflict of Interest}
The authors declare that they have no known competing financial interests or personal relationships that could have appeared to influence the work reported in this paper.

\section*{Author Contributions}
\textbf{Yongzheng Zhu:} Conceptualization; Data curation; Formal analysis; Investigation; Methodology; Resources; Software; Validation; Visualization; Writing – original draft; Writing – review \verb+&+ editing.
\textbf{Weizheng Chen:} Conceptualization; Resources; Writing – review \verb+&+ editing.
\textbf{Jian Deng:} Conceptualization; Resources; Writing – review \verb+&+ editing.
\textbf{Xin Bian:} Conceptualization; Formal analysis; Funding acquisition; Investigation; Project administration; Resources; Supervision; Writing – review \verb+&+ editing.

\section*{Acknowledgments}
Support from the grant of the National Key R\&D Program of China under contract number 2022YFA1203200 and the National Natural Science Foundation of China with No. 12172330 is gratefully acknowledged. 
We are grateful to Mr. Shiji Zhao for his assistance with the data analysis.



\bibliographystyle{unsrt} 
\bibliography{main} 

\end{multicols}

\makeentitle

\end{document}